\documentclass[journal=jacsat,manuscript=article]{achemso}

\usepackage{amsmath}
\usepackage{amssymb}
\usepackage{graphicx}
\usepackage{babel}
\usepackage{array}
\usepackage{verbatim}
\usepackage[colorlinks=true, pdfstartview=FitV, linkcolor=black, citecolor=black, urlcolor=black]{hyperref} % enable links black not blue

\usepackage{textcomp}

\def\H        {{$^1$H \/}}

\def\F        {{$^{19}$F \/}}

\def\cf       {{\it cf. \/}}

\def\BFO      {{BiFeO$_3$}}
\def\CFB	  {{CoFeB}}

\newcommand{\mr}[1]{\mathrm{#1}}
\newcommand{\unit}[1]{\,\mathrm{#1}}
\newcommand{\um}{\,\mu{\rm m}}

\newcommand{\Ares}{A_\mr{res}}
\newcommand{\Aset}{A_\mr{set}}
\newcommand{\BNV}{B_\mr{NV}}
\newcommand{\Brms}{B_\mr{rms}}
\newcommand{\daAM}{d_1^\mr{AM}}
\newcommand{\daFM}{d_1^\mr{FM}}
\newcommand{\dbAM}{d_2^\mr{AM}}
\newcommand{\dbFM}{d_2^\mr{FM}}

\newcommand{\fres}{f_\mr{res}}
\newcommand{\fset}{f_\mr{set}}
\newcommand{\Dfres}{\Delta f_\mr{res}}

\author{
	Zhewen~Xu$^{1,2}$,
	Marius~L.~Palm$^{1}$, %Marius~L.~Palm$^{1}$,
	William~S.~Huxter$^{1,\dagger}$, %William~S.~Huxter$^{1}$,
	Konstantin~Herb$^{1}$,
	John~M.~Abendroth$^{1}$,
	Karim~Bouzehouane$^{3}$,
	Olivier~Boulle$^{4}$,
	Mihai~S.~Gabor$^{5}$,
	Joseba~Urrestarazu~Larranaga$^{4}$,
	Andrea~Morales$^{2}$,
	Jan~Rhensius$^{2}$,
	Gabriel~F.~Puebla-Hellmann$^{2}$, %Gabriel~F.~Puebla-Hellmann$^{2}$,
	and Christian~L.~Degen$^{1,6}$
}
\affiliation{
	$^1$Department of Physics, ETH Z\"urich, Otto Stern Weg 1, 8093 Z\"urich, Switzerland.
	$^2$QZabre AG, Neubrunnenstr. 50, 8050 Z\"urich, Switzerland.
	$^3$Laboratoire Albert Fert, CNRS, Thales, Universit\'e Paris-Saclay, 91767 Palaiseau, France.
	$^4$Universit\'e Grenoble Alpes, CNRS, CEA, SPINTEC, 38054 Grenoble, France.
	$^5$Technical University of Cluj-Napoca, Memorandumului 28, Cluj-Napoca 400347, Romania.
	$^6$Quantum Center, ETH Z\"urich, 8093 Z\"urich, Switzerland.
	$^\dagger$Present address: Department of Physics, Massachusetts Institute of Technology, Cambridge, Massachusetts 02139, USA
}
\email{degenc@ethz.ch}

\title{Minimizing sensor-sample distances in scanning nitrogen-vacancy magnetometry}

\begin{document}
	
\begin{center}\date{\today}\end{center}

%%% Abstract
\clearpage
\section{Abstract}

Scanning magnetometry with nitrogen-vacancy (NV) centers in diamond has led to significant advances in the sensitive imaging of magnetic systems.  The spatial resolution of the technique, however, remains limited to tens to hundreds of nanometers, even for probes where NV centers are engineered within 10\,nm from the tip apex.
Here, we present a correlated investigation of the crucial parameters that determine the spatial resolution: the mechanical and magnetic stand-off distances, as well as the sub-surface NV center depth in diamond.  We study their contributions using mechanical approach curves, photoluminescence measurements, magnetometry scans, and nuclear magnetic resonance (NMR) spectroscopy of surface adsorbates.
We first show that the stand-off distance is mainly limited by features on the surface of the diamond tip, hindering mechanical access.  Next, we demonstrate that frequency-modulated atomic force microscopy (FM-AFM) feedback partially overcomes this issue, leading to closer and more consistent magnetic stand-off distances ($26-87\unit{nm}$) compared to the more common amplitude-modulated (AM-AFM) feedback ($43-128\unit{nm}$).
FM operation thus permits improved magnetic imaging of sub-100-nm spin textures, shown for the spin cycloid in \BFO\ and domain walls in a CoFeB synthetic antiferromagnet.  Finally, by examining \H and \F NMR signals in soft contact with a polytetrafluoroethylene surface, we demonstrate a minimum NV-to-sample distance of 7.9$\pm$0.4\,nm.

\section{Keywords}
Diamond NV center, Scanning probe microscopy, Spatial resolution, Magnetic imaging, Capillary bridge, Surface adsorbates, NMR spectroscopy

%%% Introduction
%\section{Introduction}
\clearpage

Magnetic imaging techniques are indispensable tools in nanoscale research, with applications ranging from fundamental research in condensed matter and materials physics to metrology and device characterization in the engineering disciplines.  For example, magnetic force microscopy~\cite{rugar1990magnetic, hartmann1999magnetic} and Lorentz transmission electron microscopy~\cite{marshall1999lorentz, yu2010real} are widely used to reveal domain patterns and spin textures, which are essential for understanding the energy scales and dynamics of magnetically ordered systems.  Similar and complementary information is available from a range of other techniques, such as scanning superconducting quantum interference device microscopy~\cite{kirtley1999scanning}, a number of scanning tunneling microscopies~\cite{choi2019electronic, nuckolls2020strongly, zhang2022promotion}, and X-ray methods such as magnetic circular dichroism and magnetic linear dichroism with photo-emission electron microscopy~\cite{grzybowski17}.
While all techniques offer nanoscale or even atomic spatial resolution, each method has its specific requirements that limit applications to certain samples or operating conditions.  For example, some techniques only accept samples as thin films, while others require atomically flat and conducting surfaces, an ultra-high vacuum environment, or cryogenic operation.  In addition, X-ray investigations rely on access to large-scale facilities where measurement time is limited.

Scanning magnetometers based on nitrogen-vacancy centers (SNVMs) are an important recent addition to the set of nanoscale magnetic imaging instruments~\cite{chernobrod05,degen08apl,balasubramanian08}.  SNVMs exploit a single defect spin in a diamond scanning tip as an atomic-size magnetic sensor, allowing for quantitative and sensitive stray field imaging at sub-100-nm spatial resolution.  Such microscopes are table-top instruments, operate under ambient conditions, and are compatible with a wide range of samples, including bulk and thin-film materials.
Initially applied to study magnetic textures in ferromagnets~\cite{rondin13,tetienne14,tetienne15,dussaux16}, the technique has since been extended to antiferromagnets~\cite{appel19, wornle19, wornle21, hedrich21, finco21}, multiferroics~\cite{gross17, chauleau20, lorenzelli21}, two-dimensional ferromagnets~\cite{thiel19, sun21, fabre21}, skyrmions~\cite{dovzhenko18, gross18, jenkins19}, superconducting vortices~\cite{thiel16, pelliccione16}, and nanoscale current distributions~\cite{chang17, palm24}.
While SNVM offers excellent sensitivity in real space imaging, the spatial resolution, typically tens of nanometers, is modest compared to related techniques like magnetic force microscopy (MFM) or scanning tunneling microscopy (STM)~\cite{schmid10}.  The limited spatial resolution hinders investigation of a range of interesting phenomena on the $\sim 10\unit{nm}$ lengthscale, such as the internal structure of magnetic domain walls~\cite{bode2006atomic}, superlattice structures~\cite{zhang2017interlayer} or magnetic signatures related to wave effects of electrons~\cite{crommie1993confinement,heller1994scattering}.

In SNVM, the spatial resolution is directly set by the vertical separation (stand-off) between the atomic-size magnetic sensor and the sample surface~\cite{chang17}. Therefore, to attain a high spatial resolution, the stand-off must be reduced as much as possible.  Early approaches based on grafting a diamond nanoparticle to a commercial AFM probe occasionally reported excellent stand-off distances ($15-25\unit{nm}$~\cite{tisler2013single,chang17}).  However, this approach is difficult to reproduce and the quantum properties of NV centers in nanocrystals are generally poor~\cite{rondin14,schirhagl14}.  By contrast, state-of-the-art scanning probes fabricated by top-down lithography from single-crystalline diamond substrates~\cite{maletinsky12,zhu23} offer excellent NV properties.  However, stand-off distances ($35-120\unit{nm}$~\cite{hingant2015measuring, tetienne2015nature, gross2016direct, gross17, rohner2019111, zhong2022quantitative, finco2022imaging,finco2023single, pham24}) are generally larger and vary greatly.  The reasons for such large stand-offs are unclear, as the nominal sub-surface depth of NV centers is expected to be around $10\unit{nm}$, based on the nitrogen implantation energy during NV synthesis.  Possible explanations are tip topography caused by surface roughness or particle pickup, tip tilt relative to the sample, meniscus formation by surface adsorbates between tip and sample, or a biased selection of deep NV centers in tip fabrication.  Routinely achieving standoff distances below $50\unit{nm}$ (ideally less than $10\unit{nm}$) remains a fundamental and practical challenge.

%%%%%

In this work, we present a systematic study of stand-off distances in SNVM imaging.  First, we characterize the critical vertical distance parameters -- the mechanical stand-off ($d_1$), the magnetic stand-off ($d_2$), and the sub-surface NV depth ($d_3$), \cf~Fig.~\ref{fig1}a -- to form a detailed picture of the tip-sample interaction.
We show that there is considerable variation in magnetic stand-off between probes, likely due to a combination of inadvertent tip contamination and meniscus formation common to large-diameter AFM probes.  By contrast, the stand-off contribution due to the sub-surface depth of NV centers is minor.  We further show that using frequency-modulation (FM) feedback for AFM tip approach and optimizing control parameters, the magnetic stand-off can be lowered by $\sim 16\unit{nm}$ (median) compared to amplitude-modulation (AM) feedback.
We demonstrate measurement improvement by imaging \BFO\ and \CFB\ magnetic test samples at stand-offs down to $24.3\pm 4.6\unit{nm}$, as well as NMR detection of surface films in soft contact with a stand-off of $7.9\pm 0.4\unit{nm}$.

\section{Results and Discussion}

\subsection{Diamond NV tips}

\begin{figure}[!tb]
	\includegraphics[width=0.90\columnwidth]{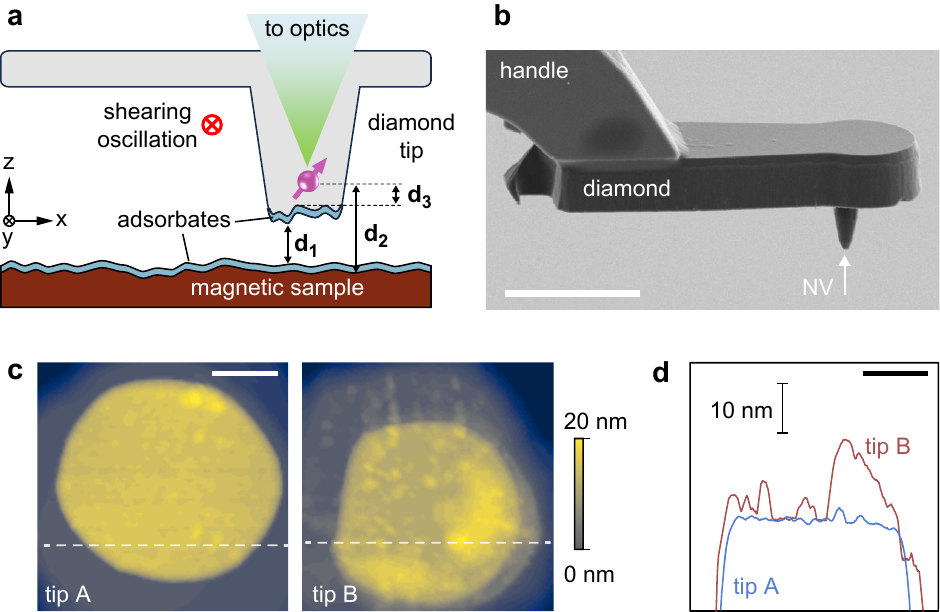}
	\caption{
		\textbf{Scanning NV microscope and diamond tip.}
		\textbf{a.} Schematic of scanning probe and surface, indicating the mechanical stand-off ($d_1$), magnetic stand-off ($d_2$), and sub-surface depth of the NV center ($d_3$).  The diamond probe is oscillated along $y$ (shear mode) using a tuning-fork actuator.  A combination of optical and microwave excitation is used to control and read out the NV spin (pink).
		\textbf{b.} Electron micrograph of a diamond scanning probe. The NV center is located at the tip apex (arrow).  Scale bar, $10\unit{\um}$.
		\textbf{c.} AFM topography of the tip apex from two representative diamond probes with smooth (tip A) and rough (tip B) surfaces, respectively.
		\textbf{d.} Line cuts along the dashed lines in {\bf c}. Scale bars in {\bf c,d} are $200\unit{nm}$.
	}
	\label{fig1}
\end{figure}

Our diamond probes~\cite{qzabre} are fabricated from single-crystal diamond substrates using a series of e-beam lithography and dry etching steps~\cite{zhu23}. Each probe contains a single NV defect center approximately 10\,nm from the tip apex formed by low-energy nitrogen implantation ($7\unit{keV}$ $^{15}$N$^+$) and subsequent annealing~\cite{janitz22}.  Tips have a tapered geometry with top diameters between $350-400\unit{nm}$~\cite{zhu23} which protrude from a larger diamond piece that is glued, via a cantilever handle, to a quartz tuning fork anchored on a ceramic chip for easy handling and electrical contacting (Fig.~\ref{fig1}b).  All measurements are performed with a commercial scanning NV magnetometer~\cite{qzabre} under ambient conditions, \cf Fig.~\ref{fig1}a and Methods.

Figs.~\ref{fig1}c,d shows AFM scans of the tip apex from two diamond probes.
These tips are from a separate chip~\cite{zhu23} and have larger diameters ($\sim 700\unit{nm}$) compared to the scanning tips analyzed in the rest of this study.
Although the tip apex is in general very smooth (tip~A in Fig.~\ref{fig1}c), with an rms roughness below one nanometer, some tips (tip~B) showed larger features with typical peak heights of $20-50\unit{nm}$ (see Supplementary Note~1 for statistics).
These features are important, as they hinder mechanical access during tip approach and therefore increase the stand-off distance.
The likely origin for features seen on tips of type~B are residues from tip lithography, such as mask material or photoresist.  The images in Fig.~\ref{fig1}c are taken on pristine tips; images acquired from scanning probes in reverse AFM mode before use (Supplementary Note~1) revealed similar features on the tip surface.  As with any AFM imaging, tips tend to accumulate more material during use when scanning in close contact~\cite{lo1999} and under prolonged laser exposure~\cite{parthasarathy2024role} even under UHV conditions~\cite{enachescu2004}.

\subsection{Mechanical approach curves}

\begin{figure}[!tb]
	\includegraphics[width=0.65\columnwidth]{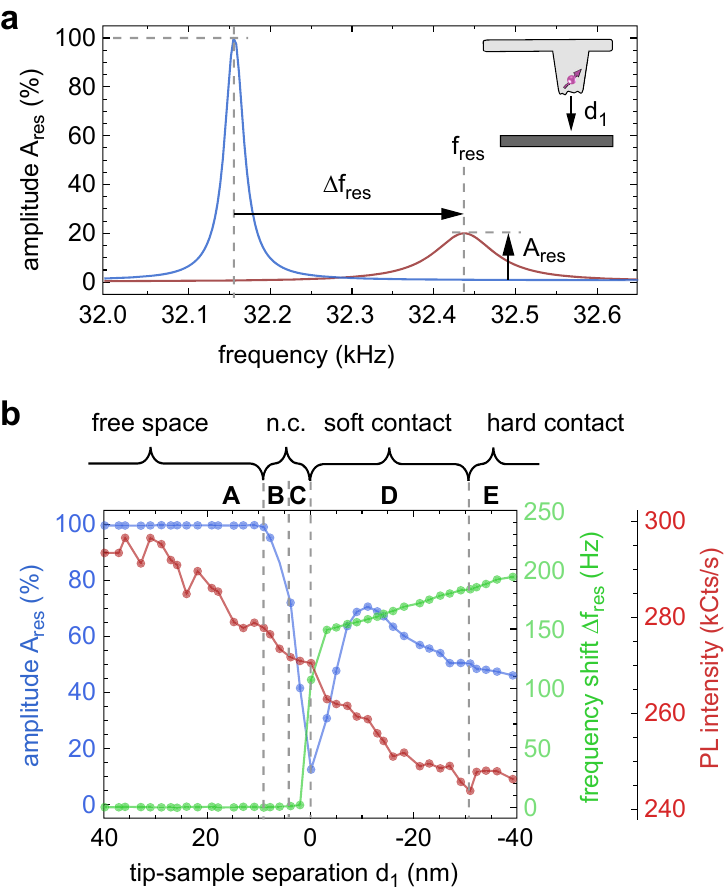}
	\caption{
		{\bf Tuning fork control and approach curves.}
		\textbf{a.} Tuning fork resonance curves in free space (blue) and in hard contact (red). $\Ares$ is the resonant amplitude and $\Dfres$ is the frequency shift, given relative to the free-space ($d_1\rightarrow\infty$) value. 
		\textbf{b.} Approach curves plotting $\Ares$ (blue), $\Dfres$ (green) and the PL intensity (red) as a function of $d_1$.  The contact point $d_1=0$ is defined by the minimum in $\Ares$.  \textbf{A}-\textbf{E} identify different interaction regimes discussed in the text.  The data shown are for diamond tip NV14 and a Si/SiO$_2$ sample surface.   n.c. = non-contact.
	}		
	\label{fig2}
\end{figure}

We begin our study by quantifying the mechanical stand-off distance, $d_1$, corresponding to the physical separation between the lowest point of the diamond probe tip and the highest point of the sample surface (see Fig.~\ref{fig1}a).  For this purpose we record the resonant amplitude ($\Ares$) and frequency ($\fres$) of the tuning fork (Fig.~\ref{fig2}a) while slowly approaching the sample surface.  By plotting $\Ares$ and $\fres$ as a function of $d_1$, we obtain the approach curves shown in Fig.~\ref{fig2}b.  In addition, we monitor the photo-luminescence (PL) intensity of the NV center under constant green laser illumination.  Since large-diameter tips operated in shear mode behave differently from standard tapping mode AFM tips~\cite{hoppe2005}, and because the understanding of the approach curves is crucial for choosing the set-point and minimizing $d_1$, we now discuss Fig.~\ref{fig2}b in some detail.

In the free-space regime, {\bf A}, no mechanical tip-sample interaction is observed. $\Ares$ and $\fres$ take their free-space values. However, a reduction in the PL intensity is already visible.  The reduction in PL is mainly due to an optical cavity forming between the parallel tip and sample surfaces~\cite{buchler2005measuring, israelsen2014increasing, ernst2019planar}.  In region {\bf B}, the amplitude $\Ares$ starts decreasing due to dissipative tip-sample interactions~\cite{drummond1996measurement, pfeiffer2002lateral, karrai2000interfacial} while $\fres$ is largely unaffected, indicating that dissipative tip-sample interactions occur prior to conservative tip-sample interactions.  In region {\bf C}, conservative tip-sample interactions start contributing, leading to a rise in $\fres$ and a further reduction in $\Ares$.  Crossing from region {\bf C} into region {\bf D}, $\Ares$ assumes a sharp minimum.  Since this minimum is a well-defined feature, we use it to define the ``contact point'' ($d_1=0$) in the approach curve~\cite{gottlich2000noncontact, pfeiffer2002lateral}.  The contact point reflects a soft contact, and the tip can be approched further towards the sample.
In region {\bf D}, the amplitude increases again and eventually reaches a local maximum.  This maximum is attributed to adsorbates, such as water in ambient conditions, filling the gap between the diamond tip and the sample surface. As a consequence, the cantilever experiences an increasing shear force and begins bending elastically~\cite{karrai2000interfacial}.  Finally, in region {\bf E}, the PL signal saturates and $\Ares$ and $\fres$ vary slowly with the $z$-position, indicating that a hard contact has been made.

To control the tip-sample distance, we feedback on either the amplitude or frequency signal (Methods).  In amplitude-modulated (AM) feedback mode, the tuning fork is driven at a fixed frequency and the oscillation amplitude is held at a constant set-point by adjusting the $z$-position using a proportional-integral (PI) controller. The set-point may be arbitrarily chosen, typically between $20-90\%$ of the free amplitude (\cf Fig.~\ref{fig2}a).  In frequency-modulated (FM) feedback mode, the tuning fork is driven at resonance with the help of a phase-locked loop (PLL), and the frequency shift (typically $5-200\unit{Hz}$) is held at a constant set-point.
%The mechanical stand-off $d_1$ is then defined as the relative distance between the set-point and the contact point.

We find that the FM feedback has several key advantages over the AM feedback:  first, because the $\Ares$ signal reacts earlier than the $\fres$ signal, the AM set-point is typically further from the sample than the FM set-point.
Further, the non-monotonic behavior of $\Ares$, especially the dip and local maximum in regions {\bf C} and {\bf D}, tends to make the feedback unstable.  As a consequence, a set-point in regions {\bf B-C} must be chosen in AM mode.
By contrast, $\fres$ increases monotonically with decreasing $d_1$ even deep in the soft-contact regime, allowing for a much closer approach combined with a more robust and reproducible feedback in FM mode.
Therefore, with a careful tuning of $\fset$, the FM mode is expected to allow for substantially smaller and more consistent stand-off distance compared to the AM mode.

\subsection{Magnetic stand-off distance}

Next, we investigate the magnetic stand-off, $d_2$, given by the vertical distance between the NV center and the top surface of the magnetic sample (Fig.~\ref{fig1}a).  Our calibration sample is an out-of-plane (OOP) magnetized ferromagnetic Co film of $1.6\unit{nm}$ thickness that is lithographically patterned into a two-micron-wide stripe~\cite{wornle2021nanoscale, luo2019chirally}.
% (Pt[6.0\,nm]/Co[1.6\,nm]/AlO$_x$[2.0\,nm])
%
To determine the magnetic stand-off, we take magnetometry line scans across the stripe and fit the data to an analytical function describing the stray field from a uniformly magnetized OOP stripe~\cite{hingant2015measuring,tetienne2015nature} (Methods).

\begin{figure*}[!tb]
	\includegraphics[width=1.00\textwidth]{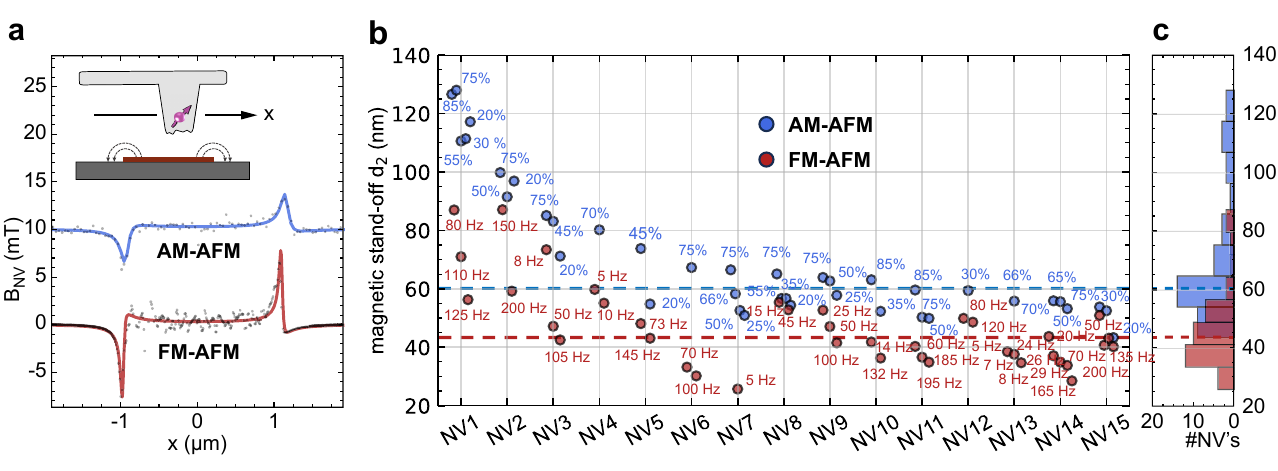}
	\caption{
		\textbf{Magnetic stand-off distance.}
		\textbf{a.} Stray field scans for NV7 obtained by scanning the tip across an OOP-magnetized ferromagnetic stripe (inset).  The upper trace (blue, offset by $10\unit{mT}$ for clarity) is recorded in AM mode and the lower trace (red) in FM mode. Dots show the data and solid lines are the analytical fits to the stray field profile (Methods). 
		\textbf{b.} Scatter plot of magnetic stand-off distances $d_2$ from diamond probes (NV1-NV15), totaling 77 scans.  NV indices are ordered according their AM-AFM stand-off.  Blue and red numbers indicate set-point parameters (see text).
		\textbf{c.} Histogram of the data shown in \textbf{b}. Dashed horizontal lines are median values.
	}
	\label{fig3}
\end{figure*}

Fig.~\ref{fig3}a depicts measured stray field profiles and model fits for one representative tip (NV7) acquired in AM (blue) and FM (red) feedback mode.  The FM feedback yields a sharper field profile and a $\sim 2\times$ larger absolute stray field, as expected for a scan in closer proximity.  For this example, the fitted magnetic stand-off distances are $58.3\pm1.5\unit{nm}$ (AM) and $25.7\pm1.4\unit{nm}$ (FM), respectively.  This result is consistent with the $\propto d_2^{-1}$ scaling of the stray field peak from a step edge (Methods).
Fig.~\ref{fig3}b presents the data from fifteen diamond probes (see Supplementary Note~3 for full dataset).  We find $d_2$ to show large variations between probes and set-point parameters, with values ranging from $43-128\unit{nm}$ in AM mode and from $26-87\unit{nm}$ in FM mode.  We also find that FM feedback leads to a consistently lower stand-off and reduced spread in $d_2$.  A histogram analysis (Fig.~\ref{fig3}c) shows that the median stand-off value is reduced from $59.6\unit{nm}$ (AM) to $43.1\unit{nm}$ (FM), corresponding to a net reduction of $16.5\unit{nm}$.  This significant reduction has a strong impact on the image resolution and strength of the stray field signal (see below, Fig.~\ref{fig5}), offering improved characterization in scanning magnetometry applications.

\subsection{Sub-surface depth of NV centers}
%\label{sec:nvnmr}

In a third step, we determine the sub-surface depth, $d_3$, of NV centers in the diamond probe by the technique of NV-NMR~\cite{mamin13,staudacher13}.  Specifically, we use dynamical decoupling of the NV center to detect the magnetic noise from the \H spins contained in the adsorbate layer on the diamond surface~\cite{degen09,loretz14apl}.
Because the intensity of the \H NMR signal quickly decreases with depth $d_3$, the peak magnitude can be used as a reliable and quantitative depth gauge for near-surface NV centers~\cite{pham16,janitz22}.

\begin{figure}[!tb]
	\includegraphics[width=0.60\columnwidth]{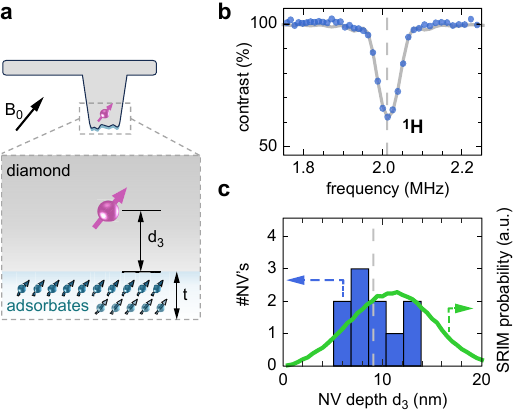}
	\caption{
		\textbf{Sub-surface depth of NV centers.}
		\textbf{a.} We calibrate the sub-surface depth $d_3$ by measuring the magnitude of the \H NMR signal from the $t\sim 1\unit{nm}$ layer of surface adsorbates~\cite{janitz22}. $B_0$ is the bias field.
		\textbf{b.} \H NMR spectrum for NV7 recorded at $B_0=47.3\unit{mT}$ using dynamical decoupling spectroscopy~\cite{bylander11}.  Dots are the experimental data and the curve is a least-squares fit to the analytical model.  Dashed line is the expected \H resonance positions at this field.
		\textbf{c.} Histogram of the measured $d_3$ from ten diamond probes. Gray dashed line is the median value.  Green curve is the result of a SRIM simulation of the $^{15}$N$^{+}$ ion distribution with depth $d_3$.  }
	\label{fig4}
\end{figure}

Fig.~\ref{fig4}b shows the \H NMR spectrum of a representative diamond probe together with the fit to an analytical model~\cite{loretz14apl} (see Supplementary Note~4 for further NMR spectra).
The depth extracted from the fit is $d_3 = 5.7 \pm 0.2\unit{nm}$ (Methods).
A histogram collecting measurements from ten diamond probes is shown in Fig.~\ref{fig4}c. We find that NV depths range from $d_3 = 5.1-13.9\unit{nm}$ with a median of $9.1 \unit{nm}$ (dashed line).  These values are in good agreement with a complementary Stopping and Range of Ions in Matter (SRIM) Monte-Carlo simulation~\cite{ziegler1985stopping} for the 7\,keV $^{15}$N$^{+}$ ions used in tip fabrication ($10.8\pm 4.0\unit{nm}$)~\cite{abendroth22} (solid curve).
From Fig.~\ref{fig4}c, we conclude that the sub-surface depths for our scanning probes are consistent with the chosen implantation energy.  In particular, we find no selection bias towards deeper NV centers, and the narrow distribution in $d_3$ cannot explain the large variation in magnetic stand-off $d_2$.

\begin{table}[!b]
		\begin{tabular}{c|rrr|rrr|r}
			\hline\hline
			diamond & \multicolumn{3}{c}{mechanical stand-off} & \multicolumn{3}{c}{magnetic stand-off} & \multicolumn{1}{c}{NV depth} \\
			probe & $\daAM$ & $\daFM$ & $\daAM-\daFM$ & $\dbAM$ & $\dbFM$ & $\dbAM-\dbFM$ & \multicolumn{1}{c}{$d_3$} \\ \hline
   			NV2 & 2.5\,nm & -25.5\,nm & 28.0\,nm & 99.8\,nm & 59.2\,nm & 40.6\,nm & 12.9\,nm  \\
			NV15 & 3.5\,nm &  -2.5\,nm & 6.0\,nm  & 53.8\,nm & 43.1\,nm & 10.7\,nm & 5.1\,nm  \\
			\hline\hline
		\end{tabular}
	\caption{
		\label{table1}
		Mechanical stand-off ($d_1$), magnetic stand-off ($d_2$) and sub-surface NV depth ($d_3$) for two diamond probes. Setpoint parameters are: $\Aset=75\%$ and $\Dfres=200\unit{Hz}$ (NV2), $\Aset=75\%$ and $\Dfres=135\unit{Hz}$ (NV15).
		Note that $d_1 + d_3 \neq d_2$ due to our definition of the $d_1=0$ mechanical contact point (see Fig.~\ref{fig2}b).
	}
\end{table}
Table~\ref{table1} concludes our stand-off analysis by presenting values for the distances $d_1$, $d_2$ and $d_3$ for two scanning probes.  Comparison between $\daAM-\daFM$ and $\dbAM-\dbFM$ shows that the reduction in stand-off is roughly consistent between the mechanical and magnetic measurements, although the reduction is greater for the latter.  This shows that our feedback control is reliable in adjusting the stand-off and does not lead to unexpected variation.  Comparison with $d_3$ also confirms that the sub-surface NV depth is only a minor contribution to the magnetic stand-off $d_2$. 

Table~\ref{table1} also exemplifies that the magnetic stand-off varies greatly between probes, corroborating our findings from Fig.~\ref{fig3}.
We hypothesize that the large variations are due to a combination of topographic features on the diamond tip (see AFM images in Fig.~\ref{fig1}c,d) and irregular meniscus formation~\cite{weeks2005} as the tip moves into soft contact with the sample surface; in the future, these could be reduced by tip cleaning protocols~\cite{gan2009} and operation under controlled atmosphere.
FM feedback partially surmounts the mechanical barrier, placing the tip deep into the soft contact regime.  Because the tuning fork has a high stiffness, a considerable force can be applied to the tip.  This force will tend to level out soft features on both surfaces and reduce tip tilt, if present.
Below, we provide further evidence for this hypothesis by detecting the \F NMR signal from a PTFE film in soft contact.

\subsection{Magnetic imaging with high spatial resolution}

\begin{figure*}[!tb]
\includegraphics[width=0.8\columnwidth]{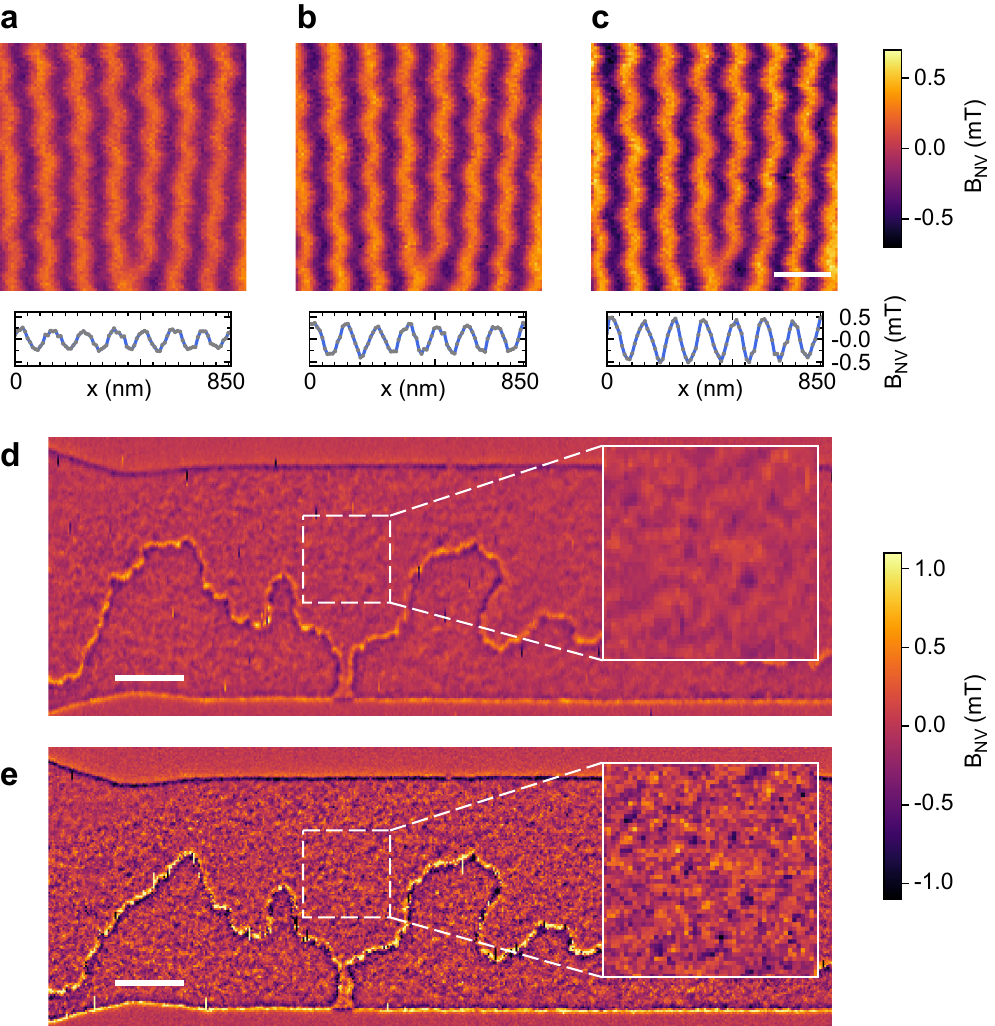}
\caption{
	\textbf{High-resolution magnetic field imaging of \BFO\ and \CFB.}
	\textbf{a-c.} Quantitative stray field maps of a \BFO\ thin film.
	\textbf{a.} AM-AFM at 50\% set-point.
	\textbf{b.} FM-AFM at $\Dfres = 65\unit{Hz}$ set-point.
	\textbf{c.} FM-AFM at $\Dfres = 155\unit{Hz}$ set-point.
	The line cuts underneath the images show the periodic stray field oscillation of the \BFO\ spin cycloid.  Stray field amplitudes (peak-to-peak) are $0.47\pm 0.03\unit{mT}$ for {\bf a}, $0.67\pm 0.07\unit{mT}$ for {\bf b}, and $0.87\pm 0.05\unit{mT}$ for {\bf c}.
	Scale bar, $200\unit{nm}$.
	\textbf{d,e.} Quantitative stray field maps of a \CFB\ synthetic antiferromagnetic racetrack measured in AM mode ({\bf d}, 25\% set-point) and FM mode ({\bf e}, $\Dfres = 110\unit{Hz}$ set-point).
	The dashed-squared regions are shown at higher magnification as insets.
	Scale bars are $1\unit{\mu m}$ and insets are $1.25\times 1.25\unit{\um}$, respectively.
	}
\label{fig5}
\end{figure*}

We now demonstrate that the lower stand-off available through FM feedback allows for a much improved spatial resolution in nanoscale imaging magnetometry applications.  As our first example, we present images of the spin cycloid of bismuth ferrite (\BFO), an archetypal multiferroic material with non-collinear antiferromagnetic textures~\cite{gross17,chauleau2020electric,haykal2020antiferromagnetic,zhong2022quantitative,finco2022imaging, dufour2023onset} (see Methods and Supplementary Note~5 for further details).
Figs.~\ref{fig5}a-c show scanning NV images recorded in AM mode (panel a) and in FM mode with weak (panel~b) and strong (panel~c) frequency set-points.  The line cuts plotted underneath the images reveal that the peak-to-peak value of the stray field increases by 85\% when going from AM (panel~a) to FM (panel~c) feedback.

Because the stray field of a periodic magnetic structure decays as $e^{-2\pi d_2/\lambda}$ with $d_2$, where $\lambda$ is one full spatial period, we can relate the signal increase to the change in stand-off distance.  Considering a period of $\lambda = 119\unit{nm}$ (extracted using Fourier analysis of Fig.~\ref{fig5}c) and the 85\% signal change, the reduction in stand-off is approximately $12\unit{nm}$.  This value is consistent with the result obtained using inverse Fourier filtering (Supplementary Note~5).  Overall, Figs.~\ref{fig5}a-c demonstrate that even a modest reduction in the stand-off distance can lead to large improvements in the magnetic signal.

Figs.~\ref{fig5}d,e present a second example of comparative imaging using a fully compensated synthetic antiferromagnetic racetrack sample.  The sample is composed of two $10.25\unit{\AA}$-thick layers of CoFeB separated by a Ru/Pt spacer leading to out-of-plane antiferromagnetic coupling (Methods).
% exact layering: Ta(30)/Pt(40)/CoFeB(10.25)/Ru(7.2)/Pt(4.5)/CoFeB(10.25)/Ru(5)/Ta(15).
% capping layer (Ru/Ta) is 2nm
Clearly, the FM image (panel e) shows the device outline and domain walls within the racetrack in crisper detail compared to the AM image (panel d).  The insets reveal that the root-mean-square (rms) values of the stray field increase from $0.13\unit{mT}$ (AM) to $0.30\unit{mT}$ (FM), corresponding to a signal rise by 130\%.  Separate fits to the stray field profiles at the edge of the racetrack (Supplementary Note~6) show that the magnetic stand-off is reduced from $d_2=43.6\pm 8.2\unit{nm}$ (AM) to $d_2=24.3\pm 4.6\unit{nm}$ (FM).  The latter value is the smallest magnetic stand-off distance measured in this study.

\subsection{Detection of meniscus formation and molecular uptake}

\begin{figure}
	\includegraphics[width=0.55\columnwidth]{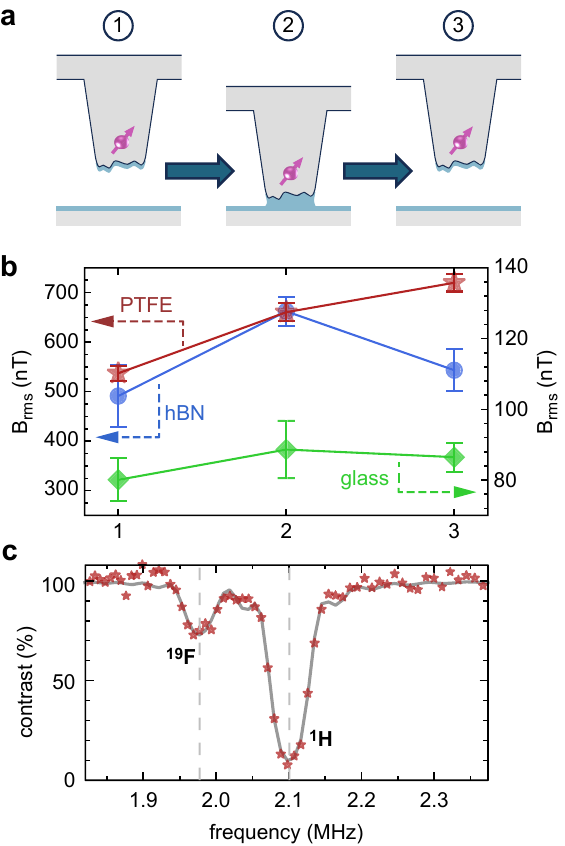}
	\caption{
		\textbf{Investigation of capillary bridge formation and molecular adsorption using NV-NMR.}
		\textbf{a.} Experimental sequence: measurements are performed in free-space (1), soft contact (2), and after retraction to free space (3).
		\textbf{b.} Magnitude of the \H NMR peak ($\Brms$) as a function of position 1-2-3.
		Data shown are for a hBN (blue, acquired with NV15), PTFE (red, NV7) and glass substrate (green, NV16).
		Note that the data were taken with different tips (different $d_3$), such that $\Brms$ values are not directly comparable between substrates.
		\textbf{c.} NMR spectrum taken in soft contact with a PTFE surface, revealing both \F and \H signals.  The spectrum is acquired using NV7 and a bias field of $49.3\unit{mT}$.  Dashed lines are the expected resonance positions at this field.
		Fitting of the \F peak (gray curve) yields $\Brms = 295.6 \pm 21.2 \unit{nT}$, corresponding to a NV-to-PTFE distance of $d_2=7.9\pm 0.4\unit{nm}$ (Methods).
		}
	\label{fig6}
\end{figure}

Finally, in an attempt to reduce the stand-off even further, we investigate meniscus formation~\cite{weeks2005} and molecular uptake in the soft contact regime ($d_1<0$) using \H and \F NMR.  As shown in Fig.~\ref{fig6}a, we approach the tip from free space (1) to soft contact (2), and then retract the tip to the original free-space position (3).  At each position, we record the \H NMR signal using NV-NMR (Fig.~\ref{fig4}).  By calibrating the signal magnitude ($\Brms$) in the initial free-space position, we can relate changes in the peak intensity during the approach-retract cycle to the amount of \H-containing material within the detection volume of the NV center~\cite{janitz22}.

Figs.~\ref{fig6}b shows soft-contact measurements performed over three representative substrates: hexagonal boron nitride (hBN) as an important capping layer in two-dimensional materials research, polytetrafluoroethylene (PTFE) as an extremely hydro- and oleophobic material, and a glass microscopy slide (see Supplementary Note~4 for NMR spectra).  For all substrates, we observe an increase in the \H NMR peak as the tip is approached from position 1 to position 2.  This reflects an increase in the thickness of the adsorbate layer, which we explain by meniscus formation.  When retracting the tip to position 3, however, different behavior is observed for the three substrates:  for hBN and glass, the \H NMR signal decreases, albeit not to the level prior to approaching.  The decrease is stronger for hBN and almost absent for glass.  A similar ``residue effect'' has been reported previously~\cite{rugar2015proton}, and has been ascribed to a local redistribution of adsorbate molecules.  For the extremely hydrophobic PTFE, by contrast, the \H NMR signal slightly increases because molecules preferentially adhere to the diamond surface, leading the tip to collect adsorbates from the PTFE surface.

For PTFE, we have also measured the \F NMR signal. Upon tip approach (position 2), we observe a clear \F NMR peak (Fig.~\ref{fig6}c), proving that the tip is in intimate contact with the sample.  Converting the $\Brms$ magnitude of the \F NMR signal into a stand-off distance, we determine a value of $d_2 = 7.9\pm 0.4\unit{nm}$, a value only marginally larger than the sub-surface depth $d_3 = 5.7\pm0.2\unit{nm}$ for this probe (Fig.~\ref{fig4}b).  Together, our results demonstrate that in soft contact, the gap between probe and surface can be minimized and stand-off distances below $10\unit{nm}$ achieved.  The increase in $\Brms$ after retraction of the tip (Fig.~\ref{fig6}b) also indicates that soft contact can lead to an increase in adsorbates.  Measurements over repeated engage-retract cycles, however, show no increase in the \H NMR intensity (Fig.~S6) and no increase in the magnetic stand-off (Fig.~S3), suggesting that the adsorbate uptake is a one-time effect.

\section*{Conclusions}

In summary, we present significant advances to the understanding of stand-off distances in SNVM microscopy, which are crucial to improving the spatial resolution and sensitivity of the technique.  Starting with a detailed analysis of the tip approach curve, we show that an FM feedback control of the tip position achieves a consistent improvement over an AM feedback.  The advance manifests both in improved tip approach (median stand-off of $43\unit{nm}$ for FM compared to $60\unit{nm}$ for AM modes), and more consistent stand-off values between different tips and experimental runs.  The best-effort stand-off, measured by scanning across a magnetic step edge, is 24\,nm.  We demonstrate that even modest reductions in the stand-off distance can lead to dramatic improvements in image quality and signal magnitude.  These improvement are valuable not only to the ODMR-based imaging demonstrated here, but also to AC current sensing~\cite{ku20,palm22,palm24} and dynamic imaging modalities like scanning gradiometry~\cite{huxter22,huxter23}.  Finally, we explore the soft-contact regime including capillary bridge formation and molecular uptake using NV-NMR, and show that sub-10-nm stand-off distances with SNVM can be reached.
At this point, the sub-surface NV depth $d_3$ becomes relevant.  Lowering the ion implantation energy during NV synthesis or use of nitrogen-doped capping layers have demonstrated $d_3<3\unit{nm}$~\cite{loretz14apl,sangtawesin19,janitz22,ohashi13,myers14}.  These results indicate that even lower stand-off distances, perhaps less than 5\,nm, might be feasible.

Looking forward and in pursuit of a higher spatial resolution, the apex diameter of the diamond tip is a crucial parameter:  in this study, diamond probes with rather large apex diameters of several hundred nanometers are used.  While the diameter improves photon out-coupling~\cite{momenzadeh15}, it also makes the tip more prone to topographic irregularities including residues from tip fabrication and particle pick-up during scanning.  In addition, large-diameter tips are more affected by meniscus formation and demand more careful adjustment of the tip tilt.  Advanced nanofabrication techniques should allow engineering of diamond probes with substantially smaller end diameters, below $100\unit{nm}$, without unduly compromising photon yield~\cite{hedrich20,zhu23}.
Further, by operating the SNVM in a controlled atmosphere or a high-vacuum environment, surface adsorbates could be reduced or entirely eliminated.  Alternatively, passivation of the diamond surface using atomic layer deposition~\cite{jones23} or suitable chemical groups~\cite{janitz22} might also reduce tip adhesion.

%%%%%%%%%%%%%% Methods

\section{Methods}

\subsection{Scanning NV microscope}

The experiments are performed using a commercial scanning NV magnetometer (QSM, QZabre AG), which operates under ambient conditions.  Scanning images are acquired by scanning the sample underneath the tip using a three-axis piezo stage while the tip remains stationary.  The tuning-fork oscillator is actuated in shear mode to provide a force-feedback needed for adjusting the tip-sample distance and performing AFM scanning~\cite{edwards1997fast, giessibl1998high,ruiter1998tuning}.  A top-side objective is used to both illuminate the diamond tip using a green diode laser ($\lambda = 516\unit{nm}$) and collect the resultant NV PL emission (band-pass filtered between $\lambda \sim 630-800\unit{nm}$) using a single-photon counting module.  The optics are operated in a confocal configuration to suppress luminescence background.  Microwave pulses for manipulating the spin states of the NV center are generated by a short bond wire loop passing within $30\unit{\um}$ from the diamond tip.  Either a permanent magnet or a vector electromagnet is employed for generating bias magnetic fields of up to $50\unit{mT}$.

\subsection{AM and FM feedback}

The tuning fork oscillator was controlled using an Anfatec AFM controller.  The tuning fork was driven electrically by applying a voltage between the electrodes patterned on the two prongs.  To detect the tuning fork oscillation, the current between the electrodes is using a transimpedance amplifier.  The Anfatec controller combines a digital lock-in amplifier, phase-locked loop and PI feedback for controlling amplitude and frequency set-points.  In AM feedback mode, the tuning fork is driven at the free-space resonance with a constant drive amplitude, and the tuning fork oscillation amplitude held at a constant set-point (typically between $20-90\%$ of the free-space amplitude) by feeding back on the $z$-position.  In FM feeedback mode, the tuning fork oscillation is driven at resonance using the PLL, and the tuning fork resonance frequency is held at a constant offset $\Dfres$ from the free-space frequency (typically between $5-200\unit{Hz}$) by feeding back on the $z$-position.

\subsection{Diamond tip fabrication}

Diamond tips are fabricated from electronic-grade single crystalline substrates with a \{100\} surface cut (Element6) using a series of lithography and etching steps~\cite{zhu23}.  All tips characterized in this study are commercial scanning probes (QST, QZabre AG), except for the tips shown in Fig.~\ref{fig1}c,d, which are part of a larger pillar array~\cite{zhu23}.  Counts rates of NV centers range from 150\,kCts/s to 600\,kCts/s, and the continuous-wave optically-detected magnetic resonance (cw-ODMR) contrast varies between 12\% and 26\%.

\subsection{Magnetic samples}

\noindent\textit{Co stripe} -- The calibration sample for magnetic stand-off measurements~\cite{hingant2015measuring,tetienne2015nature} is a roughly $2\unit{\um}$-wide stripe of Pt(6\,nm) / Co(1.6\,nm) / Al(2\,nm) with out-of-plane (OOP) anisotropy~\cite{wornle2021nanoscale}. The stripe is milled with Ar ions through a PMMA mask patterned by electron-beam lithography.  The top Al layer is oxidized with a gentle oxygen plasma (power, 30\,W) in an oxygen pressure of 10\,mTorr to induce perpendicular magnetic anisotropy~\cite{luo2019chirally}.

\noindent\textit{\BFO} -- The \BFO\ sample is grown by pulsed-laser deposition on SmScO$_3$(110)$_o$ (o denotes orthorhombic) substrates. The thickness of the \BFO\ film is between 33 and 54\,nm~\cite{zhong2022quantitative}.

\noindent\textit{CoFeB} -- The CoFeB synthetic antiferromagnetic racetrack is fabricated from a Ta(3\,nm) / Pt(4\,nm) / CoFeB(1.025\,nm) / Ru(0.72\,nm) / Pt(0.45\,nm) / CoFeB(1.025\,nm) / Ru(0.5\,nm) / Ta(1.5\,nm) thin film stack, where numbers in parentheses represents nominal thicknesses.  The films are deposited at room temperature by DC magnetron sputtering in an argon pressure of $1\unit{mTorr}$ on high-resistivity Si substrates.

\subsection{Substrates used for detection of meniscus formation}

\noindent\textit{PTFE} -- Commercially available PTFE plates (APSOparts) were cleaned by sonication in acetone for $10\unit{min}$ followed by isopropanol for $2\unit{min}$, and then blow-dried using N$_2$ gas.

\noindent\textit{hBN} -- The hBN substrate was prepared by exfoliation from hBN crystals (courtesy of K.~Watanabe and T.~Taniguchi) onto a silicon substrate chip with a 90-nm-thick silicon dioxide layer (dry chlorinated thermal oxide with forming gas anneal) using blue tape (Nitto ELP BT-150-E-CM), following the method described in \cite{palm22}.

\noindent\textit{Glass} -- Commercial glass cover slips (Knittel Glasbearbeitung) were used with no extra cleaning steps.

\subsection{Fitting of magnetic stand-off}

The stray field profile from a uniform, OOP magnetized stripe in the thin-film limit can be modeled by two anti-parallel bound currents flowing at either edge of the stripe~\cite{hingant2015measuring}
\begin{align}
	B_x(x, d_2) &= \frac{\mu_0M_zt}{2\pi}\left[-\frac{d_2}{\left(x-x_1\right)^2 + d_2^2} + \frac{d_2}{\left(x-x_2\right)^2 + d_2^2}\right],\\
	B_z(x, d_2) &= \frac{\mu_0M_zt}{2\pi}\left[\frac{x-x_1}{\left(x-x_1\right)^2 + d_2^2} - \frac{x-x_2}{\left(x-x_2\right)^2 + d_2^2}\right],\\
	\BNV &= B_x \sin\theta \cos\varphi + B_z \cos\theta,
	\label{eq:stripe}
\end{align}
where $M_z$ is the OOP magnetization, $t$ the thickness of the ferromagnetic layer, $x_1$ and $x_2$ are the positions of stripe edges, and $\theta$ and $\varphi$ are the polar and azimuth angles of the NV center in the laboratory (scanning) frame, respectively. By fitting the measured stray field using Eq.~(\ref{eq:stripe}), the magnetic stand-off distance $d_2$ is determined. Note that the analytical model assumes a perfect OOP magnetization that is uniform across the film.  This model may be oversimplified, as the magnetization can deviate from OOP alignment at the edges due to the demagnetizing field.  The tilting will lead to an overestimation of $d_2$.  However, as long as the edge region where the magnetization deviates from OOP is much narrower than the stand-off distance, the effect is negligible~\cite{tetienne2015nature}.

\subsection{NMR depth measurements}

%We investigate the sub-surface depth of NV centers $d_3$ by measuring the magnitude of the \H NMR signal from the adsorbate layer on the diamond probe surface.
Surface adsorbates like water or hydrocarbons are naturally present on diamond under ambient conditions and have a typical thickness of $1-2\unit{nm}$~\cite{degen09,loretz14apl,janitz22}.  This adsorbate layer has served as a reliable depth gauge in previous NV-NMR studies~\cite{loretz14apl,pham16,abendroth22,janitz22}.
We detect the \H NMR signal through the reduction in NV spin coherence using dynamical decoupling spectroscopy.  The exact measurement protocol is given in Supplementary Note~4. 
The magnitude of the coherence dip can be converted into an rms value $\Brms$ of magnetic field noise generated by the \H spins. $\Brms$ is strongly dependent on the distance between the NV center and the \H layer, and can thus be used to precisely determine $d_3$.
We use the analytical expression of Refs.~\cite{loretz14apl,abendroth22},
\begin{equation}\label{d3_Brms}
	\Brms^2 = \frac{5\mu_0^2 \hbar^2 \gamma^2 \rho}{1536\pi d^3_3} \left(1 - \frac{d_3^3}{\left(d_3 + t \right)^3} \right),
\end{equation}
where $\mu_0$ is the vacuum magnetic permeability, $\hbar$ is the reduced Planck constant, $\gamma = 2\pi \times 42.57\unit{MHz/T}$ is the proton gyromagnetic ratio, $\rho$ the proton density and $t$ the thickness of the surface adsorption layer.  We assume values of $\rho = 60\unit{(nm)^{-3}}$ typical for water or hydrocarbons and $t = 1.3\unit{nm}$ from a previous study~\cite{abendroth22}.

%%%%%%%%%%%%%% Final sections
	
\section{Associated Content}

\subsection{Data Availability Statement}
The data that support the plots within this paper and other findings of this study are available from the corresponding authors upon reasonable request.

\subsection{Supporting Information}
The Supporting Information provides details on:
AFM imaging of diamond probes,
Approach curves,
Magnetic stand-off measurements,
NV depth measurements using proton NMR, and
Imaging of B\lowercase{i}F\lowercase{e}O$_3$ and C\lowercase{o}F\lowercase{e}B.

\section{Author Information}

\subsection{Author Contributions}
Z.X., G.F.P.H. and C.L.D. conceived and designed the experiments.
Z.X. performed all measurements and analyzed the data.
M.L.P., W.S.H. and A.M. contributed to the approach curve and magnetometry measurements.
K.H. contributed to the NMR measurements.
J.M.A. and K.B. performed AFM imaging of tips.
O.B., M.G., J.U.L. (CoFeB) and K.B. (\BFO) provided the magnetic test samples.
J.R. fabricated the diamond probes.
Z.X. and C.L.D. wrote the manuscript with assistance from all other authors.

\subsection{Notes}
A.M., J.R. and G.P.H. declare being co-founders and shareholders of QZabre AG, a startup company engaged in the engineering of SNVM systems and diamond NV probes.

\section{Acknowledgments}
The authors thank Bert Voigtl\"ander, Anne-D. M\"uller, Bj\"orn Josteinsson, Simon Josephy, Tianqi Zhu, Pol Welter, Salvatore Teresi, Rodrigo Garcia, and Ilaria Di Manici for advice and discussions as well as help in sample preparation and characterization.
The project received funding from the European Union's Horizon 2020 research and innovation programme under the Marie Sklodowska-Curie grant agreement No. 955671.
The work was supported by the Swiss National Science Foundation (SNSF) under Grants No. IZRPZ0\_194970, 200020\_212051, 200021\_219386 and CRSII\_222812, and from the State Secretariat for Education, Research, and Innovation (SBFO), Project ``QMetMuFuSP'', under Grant No. UeM019-8.
M.S.G. acknowledges financial support of the MRID-CNCS-UEFISCDI, Project PN-IV-P1-PCE-2023-1548, within PNCDI IV. O.B. and J.U.L. acknowledge support of the French Agence Nationale de la Recherche, Project ANR-17-CE24-0045 (SKYLOGIC), and the DARPA TEE program through Grant MIPR HR0011831554 from the DOI. This work was partly supported by the French RENATECH network, implemented at the Upstream Technological Platform in Grenoble PTA (Grant ANR-22-PEEL-0015). K.B. acknowledges support by the French Agence Nationale de la Recherche, Project ANR-21-ESRE-0031 (Equipex e-DIAMANT) and by the Sesame Ile de France, Project EX039175 (IMAGeSPIN).

%%%%%%%%%%%%%% Bibliography
	
%\bibliography{library_zx}

\providecommand{\noopsort}[1]{}\providecommand{\singleletter}[1]{#1}%
\providecommand{\latin}[1]{#1}
\makeatletter
\providecommand{\doi}
  {\begingroup\let\do\@makeother\dospecials
  \catcode`\{=1 \catcode`\}=2 \doi@aux}
\providecommand{\doi@aux}[1]{\endgroup\texttt{#1}}
\makeatother
\providecommand*\mcitethebibliography{\thebibliography}
\csname @ifundefined\endcsname{endmcitethebibliography}
  {\let\endmcitethebibliography\endthebibliography}{}
\begin{mcitethebibliography}{95}
\providecommand*\natexlab[1]{#1}
\providecommand*\mciteSetBstSublistMode[1]{}
\providecommand*\mciteSetBstMaxWidthForm[2]{}
\providecommand*\mciteBstWouldAddEndPuncttrue
  {\def\EndOfBibitem{\unskip.}}
\providecommand*\mciteBstWouldAddEndPunctfalse
  {\let\EndOfBibitem\relax}
\providecommand*\mciteSetBstMidEndSepPunct[3]{}
\providecommand*\mciteSetBstSublistLabelBeginEnd[3]{}
\providecommand*\EndOfBibitem{}
\mciteSetBstSublistMode{f}
\mciteSetBstMaxWidthForm{subitem}{(\alph{mcitesubitemcount})}
\mciteSetBstSublistLabelBeginEnd
  {\mcitemaxwidthsubitemform\space}
  {\relax}
  {\relax}

\bibitem[Rugar \latin{et~al.}(1990)Rugar, Mamin, Guethner, Lambert, Stern,
  McFadyen, and Yogi]{rugar1990magnetic}
Rugar,~D.; Mamin,~H.; Guethner,~P.; Lambert,~S.; Stern,~J.; McFadyen,~I.;
  Yogi,~T. Magnetic force microscopy: General principles and application to
  longitudinal recording media. \emph{Journal of Applied Physics}
  \textbf{1990}, \emph{68}, 1169--1183\relax
\mciteBstWouldAddEndPuncttrue
\mciteSetBstMidEndSepPunct{\mcitedefaultmidpunct}
{\mcitedefaultendpunct}{\mcitedefaultseppunct}\relax
\EndOfBibitem
\bibitem[Hartmann(1999)]{hartmann1999magnetic}
Hartmann,~U. Magnetic force microscopy. \emph{Annual review of materials
  science} \textbf{1999}, \emph{29}, 53--87\relax
\mciteBstWouldAddEndPuncttrue
\mciteSetBstMidEndSepPunct{\mcitedefaultmidpunct}
{\mcitedefaultendpunct}{\mcitedefaultseppunct}\relax
\EndOfBibitem
\bibitem[Marshall \latin{et~al.}(1999)Marshall, Klein, Dodge, Ahn, Reiner,
  Mieville, Antagonazza, Kapitulnik, Geballe, and Beasley]{marshall1999lorentz}
Marshall,~A.; Klein,~L.; Dodge,~J.; Ahn,~C.; Reiner,~J.; Mieville,~L.;
  Antagonazza,~L.; Kapitulnik,~A.; Geballe,~T.; Beasley,~M. Lorentz
  transmission electron microscope study of ferromagnetic domain walls in SrRuO
  3: statics, dynamics, and crystal structure correlation. \emph{Journal of
  applied physics} \textbf{1999}, \emph{85}, 4131--4140\relax
\mciteBstWouldAddEndPuncttrue
\mciteSetBstMidEndSepPunct{\mcitedefaultmidpunct}
{\mcitedefaultendpunct}{\mcitedefaultseppunct}\relax
\EndOfBibitem
\bibitem[Yu \latin{et~al.}(2010)Yu, Onose, Kanazawa, Park, Han, Matsui,
  Nagaosa, and Tokura]{yu2010real}
Yu,~X.; Onose,~Y.; Kanazawa,~N.; Park,~J.~H.; Han,~J.; Matsui,~Y.; Nagaosa,~N.;
  Tokura,~Y. Real-space observation of a two-dimensional skyrmion crystal.
  \emph{Nature} \textbf{2010}, \emph{465}, 901--904\relax
\mciteBstWouldAddEndPuncttrue
\mciteSetBstMidEndSepPunct{\mcitedefaultmidpunct}
{\mcitedefaultendpunct}{\mcitedefaultseppunct}\relax
\EndOfBibitem
\bibitem[Kirtley and Wikswo~Jr(1999)Kirtley, and
  Wikswo~Jr]{kirtley1999scanning}
Kirtley,~J.~R.; Wikswo~Jr,~J.~P. Scanning SQUID microscopy. \emph{Annual Review
  of Materials Science} \textbf{1999}, \emph{29}, 117--148\relax
\mciteBstWouldAddEndPuncttrue
\mciteSetBstMidEndSepPunct{\mcitedefaultmidpunct}
{\mcitedefaultendpunct}{\mcitedefaultseppunct}\relax
\EndOfBibitem
\bibitem[Choi \latin{et~al.}(2019)Choi, Kemmer, Peng, Thomson, Arora, Polski,
  Zhang, Ren, Alicea, Refael, \latin{et~al.} others]{choi2019electronic}
Choi,~Y.; Kemmer,~J.; Peng,~Y.; Thomson,~A.; Arora,~H.; Polski,~R.; Zhang,~Y.;
  Ren,~H.; Alicea,~J.; Refael,~G.; others Electronic correlations in twisted
  bilayer graphene near the magic angle. \emph{Nature physics} \textbf{2019},
  \emph{15}, 1174--1180\relax
\mciteBstWouldAddEndPuncttrue
\mciteSetBstMidEndSepPunct{\mcitedefaultmidpunct}
{\mcitedefaultendpunct}{\mcitedefaultseppunct}\relax
\EndOfBibitem
\bibitem[Nuckolls \latin{et~al.}(2020)Nuckolls, Oh, Wong, Lian, Watanabe,
  Taniguchi, Bernevig, and Yazdani]{nuckolls2020strongly}
Nuckolls,~K.~P.; Oh,~M.; Wong,~D.; Lian,~B.; Watanabe,~K.; Taniguchi,~T.;
  Bernevig,~B.~A.; Yazdani,~A. Strongly correlated Chern insulators in
  magic-angle twisted bilayer graphene. \emph{Nature} \textbf{2020},
  \emph{588}, 610--615\relax
\mciteBstWouldAddEndPuncttrue
\mciteSetBstMidEndSepPunct{\mcitedefaultmidpunct}
{\mcitedefaultendpunct}{\mcitedefaultseppunct}\relax
\EndOfBibitem
\bibitem[Zhang \latin{et~al.}(2022)Zhang, Polski, Lewandowski, Thomson, Peng,
  Choi, Kim, Watanabe, Taniguchi, Alicea, \latin{et~al.}
  others]{zhang2022promotion}
Zhang,~Y.; Polski,~R.; Lewandowski,~C.; Thomson,~A.; Peng,~Y.; Choi,~Y.;
  Kim,~H.; Watanabe,~K.; Taniguchi,~T.; Alicea,~J.; others Promotion of
  superconductivity in magic-angle graphene multilayers. \emph{Science}
  \textbf{2022}, \emph{377}, 1538--1543\relax
\mciteBstWouldAddEndPuncttrue
\mciteSetBstMidEndSepPunct{\mcitedefaultmidpunct}
{\mcitedefaultendpunct}{\mcitedefaultseppunct}\relax
\EndOfBibitem
\bibitem[Grzybowski \latin{et~al.}(2017)Grzybowski, Wadley, Edmonds, Beardsley,
  Hills, Campion, Gallagher, Chauhan, Novak, Jungwirth, Maccherozzi, and
  Dhesi]{grzybowski17}
Grzybowski,~M.~J.; Wadley,~P.; Edmonds,~K.~W.; Beardsley,~R.; Hills,~V.;
  Campion,~R.~P.; Gallagher,~B.~L.; Chauhan,~J.~S.; Novak,~V.; Jungwirth,~T.;
  Maccherozzi,~F.; Dhesi,~S.~S. Imaging current-induced switching of
  antiferromagnetic domains in cumnas. \emph{Phys. Rev. Lett.} \textbf{2017},
  \emph{118}, 057701\relax
\mciteBstWouldAddEndPuncttrue
\mciteSetBstMidEndSepPunct{\mcitedefaultmidpunct}
{\mcitedefaultendpunct}{\mcitedefaultseppunct}\relax
\EndOfBibitem
\bibitem[Chernobrod and Berman(2005)Chernobrod, and Berman]{chernobrod05}
Chernobrod,~B.~M.; Berman,~G.~P. Spin microscope based on optically detected
  magnetic resonance. \emph{J. Appl. Phys.} \textbf{2005}, \emph{97},
  014903\relax
\mciteBstWouldAddEndPuncttrue
\mciteSetBstMidEndSepPunct{\mcitedefaultmidpunct}
{\mcitedefaultendpunct}{\mcitedefaultseppunct}\relax
\EndOfBibitem
\bibitem[Degen(2008)]{degen08apl}
Degen,~C.~L. Scanning magnetic field microscope with a diamond single-spin
  sensor. \emph{Appl. Phys. Lett.} \textbf{2008}, \emph{92}, 243111\relax
\mciteBstWouldAddEndPuncttrue
\mciteSetBstMidEndSepPunct{\mcitedefaultmidpunct}
{\mcitedefaultendpunct}{\mcitedefaultseppunct}\relax
\EndOfBibitem
\bibitem[Balasubramanian \latin{et~al.}(2008)Balasubramanian, Chan, Kolesov,
  Al-Hmoud, Tisler, Shin, Kim, Wojcik, Hemmer, Krueger, Hanke, Leitenstorfer,
  Bratschitsch, Jelezko, and Wrachtrup]{balasubramanian08}
Balasubramanian,~G.; Chan,~I.~Y.; Kolesov,~R.; Al-Hmoud,~M.; Tisler,~J.;
  Shin,~C.; Kim,~C.; Wojcik,~A.; Hemmer,~P.~R.; Krueger,~A.; Hanke,~T.;
  Leitenstorfer,~A.; Bratschitsch,~R.; Jelezko,~F.; Wrachtrup,~J. Nanoscale
  imaging magnetometry with diamond spins under ambient conditions.
  \emph{Nature} \textbf{2008}, \emph{455}, 648\relax
\mciteBstWouldAddEndPuncttrue
\mciteSetBstMidEndSepPunct{\mcitedefaultmidpunct}
{\mcitedefaultendpunct}{\mcitedefaultseppunct}\relax
\EndOfBibitem
\bibitem[Rondin \latin{et~al.}(2013)Rondin, Tetienne, Rohart, Thiaville,
  Hingant, Spinicelli, Roch, and Jacques]{rondin13}
Rondin,~L.; Tetienne,~J.~P.; Rohart,~S.; Thiaville,~A.; Hingant,~T.;
  Spinicelli,~P.; Roch,~J.~F.; Jacques,~V. Stray-field imaging of magnetic
  vortices with a single diamond spin. \emph{Nat. Commun.} \textbf{2013},
  \emph{4}, 2279\relax
\mciteBstWouldAddEndPuncttrue
\mciteSetBstMidEndSepPunct{\mcitedefaultmidpunct}
{\mcitedefaultendpunct}{\mcitedefaultseppunct}\relax
\EndOfBibitem
\bibitem[Tetienne \latin{et~al.}(2014)Tetienne, Hingant, Kim, Diez, Adam,
  Garcia, Roch, Rohart, Thiaville, Ravelosona, and Jacques]{tetienne14}
Tetienne,~J.~P.; Hingant,~T.; Kim,~J.; Diez,~L.~H.; Adam,~J.~P.; Garcia,~K.;
  Roch,~J.~F.; Rohart,~S.; Thiaville,~A.; Ravelosona,~D.; Jacques,~V. Nanoscale
  imaging and control of domain-wall hopping with a nitrogen-vacancy center
  microscope. \emph{Science} \textbf{2014}, \emph{344}, 1366--1369\relax
\mciteBstWouldAddEndPuncttrue
\mciteSetBstMidEndSepPunct{\mcitedefaultmidpunct}
{\mcitedefaultendpunct}{\mcitedefaultseppunct}\relax
\EndOfBibitem
\bibitem[Tetienne \latin{et~al.}(2015)Tetienne, Hingant, Martinez, Rohart,
  Thiaville, Diez, Garcia, Adam, Kim, Roch, Miron, Gaudin, Vila, Ocker,
  Ravelosona, and Jacques]{tetienne15}
Tetienne,~J.~P. \latin{et~al.}  The nature of domain walls in ultrathin
  ferromagnets revealed by scanning nanomagnetometry. \emph{Nat. Commun.}
  \textbf{2015}, \emph{6}, 6733\relax
\mciteBstWouldAddEndPuncttrue
\mciteSetBstMidEndSepPunct{\mcitedefaultmidpunct}
{\mcitedefaultendpunct}{\mcitedefaultseppunct}\relax
\EndOfBibitem
\bibitem[Dussaux \latin{et~al.}(2016)Dussaux, Schoenherr, Koumpouras, Chico,
  Chang, Lorenzelli, Kanazawa, Tokura, Garst, Bergman, Degen, and
  Meier]{dussaux16}
Dussaux,~A.; Schoenherr,~P.; Koumpouras,~K.; Chico,~J.; Chang,~K.;
  Lorenzelli,~L.; Kanazawa,~N.; Tokura,~Y.; Garst,~M.; Bergman,~A.;
  Degen,~C.~L.; Meier,~D. Local dynamics of topological magnetic defects in the
  itinerant helimagnet {{FeGe}}. \emph{Nature Communications} \textbf{2016},
  \emph{7}, 12430\relax
\mciteBstWouldAddEndPuncttrue
\mciteSetBstMidEndSepPunct{\mcitedefaultmidpunct}
{\mcitedefaultendpunct}{\mcitedefaultseppunct}\relax
\EndOfBibitem
\bibitem[Appel \latin{et~al.}(2019)Appel, Shields, Kosub, Hedrich, Hubner,
  Fassbender, Makarov, and Maletinsky]{appel19}
Appel,~P.; Shields,~B.~J.; Kosub,~T.; Hedrich,~N.; Hubner,~R.; Fassbender,~J.;
  Makarov,~D.; Maletinsky,~P. Nanomagnetism of magnetoelectric granular
  thin-film antiferromagnets. \emph{Nano Lett.} \textbf{2019}, \emph{19},
  1682--1687\relax
\mciteBstWouldAddEndPuncttrue
\mciteSetBstMidEndSepPunct{\mcitedefaultmidpunct}
{\mcitedefaultendpunct}{\mcitedefaultseppunct}\relax
\EndOfBibitem
\bibitem[Wornle \latin{et~al.}(2019)Wornle, Welter, Kaspar, Olejnik, Novak,
  Campion, Wadley, Jungwirth, Degen, and Gambardella]{wornle19}
Wornle,~M.~S.; Welter,~P.; Kaspar,~Z.; Olejnik,~K.; Novak,~V.; Campion,~R.~P.;
  Wadley,~P.; Jungwirth,~T.; Degen,~C.~L.; Gambardella,~P. Current-induced
  fragmentation of antiferromagnetic domains. \emph{arXiv:1912.05287}
  \textbf{2019}, \relax
\mciteBstWouldAddEndPunctfalse
\mciteSetBstMidEndSepPunct{\mcitedefaultmidpunct}
{}{\mcitedefaultseppunct}\relax
\EndOfBibitem
\bibitem[Wornle \latin{et~al.}(2021)Wornle, Welter, Giraldo, Lottermoser,
  Fiebig, Gambardella, and Degen]{wornle21}
Wornle,~M.~S.; Welter,~P.; Giraldo,~M.; Lottermoser,~T.; Fiebig,~M.;
  Gambardella,~P.; Degen,~C.~L. Coexistence of {Bloch} and {Ne\'el} walls in a
  collinear antiferromagnet. \emph{Phys. Rev. B} \textbf{2021}, \emph{103},
  094426\relax
\mciteBstWouldAddEndPuncttrue
\mciteSetBstMidEndSepPunct{\mcitedefaultmidpunct}
{\mcitedefaultendpunct}{\mcitedefaultseppunct}\relax
\EndOfBibitem
\bibitem[Hedrich \latin{et~al.}(2021)Hedrich, Wagner, Pylypovskyi, Shields,
  Kosub, Sheka, Makarov, and Maletinsky]{hedrich21}
Hedrich,~N.; Wagner,~K.; Pylypovskyi,~O.~V.; Shields,~B.~J.; Kosub,~T.;
  Sheka,~D.~D.; Makarov,~D.; Maletinsky,~P. Nanoscale mechanics of
  antiferromagnetic domain walls. \emph{Nature Physics} \textbf{2021},
  \emph{17}, 064007\relax
\mciteBstWouldAddEndPuncttrue
\mciteSetBstMidEndSepPunct{\mcitedefaultmidpunct}
{\mcitedefaultendpunct}{\mcitedefaultseppunct}\relax
\EndOfBibitem
\bibitem[Finco \latin{et~al.}(2021)Finco, Haykal, Tanos, Fabre, Chouaieb,
  Akhtar, Robert-Philip, Legrand, Ajejas, Bouzehouane, Reyren, Devolder, Adam,
  Kim, Cros, and Jacques]{finco21}
Finco,~A. \latin{et~al.}  Imaging non-collinear antiferromagnetic textures via
  single spin relaxometry. \emph{Nature Communications} \textbf{2021},
  \emph{12}, 767\relax
\mciteBstWouldAddEndPuncttrue
\mciteSetBstMidEndSepPunct{\mcitedefaultmidpunct}
{\mcitedefaultendpunct}{\mcitedefaultseppunct}\relax
\EndOfBibitem
\bibitem[Gross \latin{et~al.}(2017)Gross, Akhtar, Garcia, Martinez, Chouaieb,
  Garcia, Carretero, Arthelemy, Appel, Maletinsky, Kim, Chauleau, Jaouen,
  Viret, Bibes, Fusil, and Jacques]{gross17}
Gross,~I. \latin{et~al.}  Real-space imaging of non-collinear antiferromagnetic
  order with a single-spin magnetometer. \emph{Nature} \textbf{2017},
  \emph{549}, 252\relax
\mciteBstWouldAddEndPuncttrue
\mciteSetBstMidEndSepPunct{\mcitedefaultmidpunct}
{\mcitedefaultendpunct}{\mcitedefaultseppunct}\relax
\EndOfBibitem
\bibitem[Chauleau \latin{et~al.}(2020)Chauleau, Chirac, Fusil, Garcia, Akhtar,
  Tranchida, Thibaudeau, Gross, Blouzon, Finco, Bibes, Dkhil, Khalyavin,
  Manuel, Jacques, Jaouen, and Viret]{chauleau20}
Chauleau,~J. \latin{et~al.}  Electric and antiferromagnetic chiral textures at
  multiferroic domain walls. \emph{Nature Materials} \textbf{2020}, \emph{19},
  386--390\relax
\mciteBstWouldAddEndPuncttrue
\mciteSetBstMidEndSepPunct{\mcitedefaultmidpunct}
{\mcitedefaultendpunct}{\mcitedefaultseppunct}\relax
\EndOfBibitem
\bibitem[Lorenzelli(2021)]{lorenzelli21}
Lorenzelli,~L. Development of a scanning nitrogen-vacancy-center magnetometer
  for variable temperature experiments. \emph{PhD Thesis, ETH Zurich}
  \textbf{2021}, \relax
\mciteBstWouldAddEndPunctfalse
\mciteSetBstMidEndSepPunct{\mcitedefaultmidpunct}
{}{\mcitedefaultseppunct}\relax
\EndOfBibitem
\bibitem[Thiel \latin{et~al.}(2019)Thiel, Wang, Tschudin, Rohner,
  Gutierrez-lezama, Ubrig, Gibertini, Giannini, Morpurgo, and
  Maletinsky]{thiel19}
Thiel,~L.; Wang,~Z.; Tschudin,~M.~A.; Rohner,~D.; Gutierrez-lezama,~I.;
  Ubrig,~N.; Gibertini,~M.; Giannini,~E.; Morpurgo,~A.~F.; Maletinsky,~P.
  Probing magnetism in {{2D}} materials at the nanoscale with single-spin
  microscopy. \emph{Science} \textbf{2019}, \emph{364}, 973\relax
\mciteBstWouldAddEndPuncttrue
\mciteSetBstMidEndSepPunct{\mcitedefaultmidpunct}
{\mcitedefaultendpunct}{\mcitedefaultseppunct}\relax
\EndOfBibitem
\bibitem[Sun \latin{et~al.}(2021)Sun, Song, Anderson, Brunner, Forster,
  Shalomayeva, Taniguchi, Watanabe, Grafe, Stohr, Xu, and Wrachtrup]{sun21}
Sun,~Q.; Song,~T.; Anderson,~E.; Brunner,~A.; Forster,~J.; Shalomayeva,~T.;
  Taniguchi,~T.; Watanabe,~K.; Grafe,~J.; Stohr,~R.; Xu,~X.; Wrachtrup,~J.
  Magnetic domains and domain wall pinning in atomically thin {{CrBr}}$_3$
  revealed by nanoscale imaging. \emph{Nature Communications} \textbf{2021},
  \emph{12}, 1989\relax
\mciteBstWouldAddEndPuncttrue
\mciteSetBstMidEndSepPunct{\mcitedefaultmidpunct}
{\mcitedefaultendpunct}{\mcitedefaultseppunct}\relax
\EndOfBibitem
\bibitem[Fabre \latin{et~al.}(2021)Fabre, Finco, Purbawati, Hadj-Azzem,
  Rougemaille, Coraux, Philip, and Jacques]{fabre21}
Fabre,~F.; Finco,~A.; Purbawati,~A.; Hadj-Azzem,~A.; Rougemaille,~N.;
  Coraux,~J.; Philip,~I.; Jacques,~V. Characterization of room-temperature
  in-plane magnetization in thin flakes of {{CrTe}}$_2$ with a single-spin
  magnetometer. \emph{Phys. Rev. Materials} \textbf{2021}, \emph{5},
  034008\relax
\mciteBstWouldAddEndPuncttrue
\mciteSetBstMidEndSepPunct{\mcitedefaultmidpunct}
{\mcitedefaultendpunct}{\mcitedefaultseppunct}\relax
\EndOfBibitem
\bibitem[Dovzhenko \latin{et~al.}(2018)Dovzhenko, Casola, Schlotter, Zhou,
  Buttner, Walsworth, Beach, and Yacoby]{dovzhenko18}
Dovzhenko,~Y.; Casola,~F.; Schlotter,~S.; Zhou,~T.~X.; Buttner,~F.;
  Walsworth,~R.~L.; Beach,~G. S.~D.; Yacoby,~A. Magnetostatic twists in
  room-temperature skyrmions explored by nitrogen-vacancy center spin texture
  reconstruction. \emph{Nature Communications} \textbf{2018}, \emph{9},
  2712\relax
\mciteBstWouldAddEndPuncttrue
\mciteSetBstMidEndSepPunct{\mcitedefaultmidpunct}
{\mcitedefaultendpunct}{\mcitedefaultseppunct}\relax
\EndOfBibitem
\bibitem[Gross \latin{et~al.}(2018)Gross, Akhtar, Hrabec, Sampaio, Martinez,
  Chouaieb, Shields, Maletinsky, Thiaville, Rohart, and Jacques]{gross18}
Gross,~I.; Akhtar,~W.; Hrabec,~A.; Sampaio,~J.; Martinez,~L.~J.; Chouaieb,~S.;
  Shields,~B.~J.; Maletinsky,~P.; Thiaville,~A.; Rohart,~S.; Jacques,~V.
  Skyrmion morphology in ultrathin magnetic films. \emph{Phys. Rev. Materials}
  \textbf{2018}, \emph{2}, 024406\relax
\mciteBstWouldAddEndPuncttrue
\mciteSetBstMidEndSepPunct{\mcitedefaultmidpunct}
{\mcitedefaultendpunct}{\mcitedefaultseppunct}\relax
\EndOfBibitem
\bibitem[Jenkins \latin{et~al.}(2019)Jenkins, Pelliccione, Yu, Ma, Li, Wang,
  and Jayich]{jenkins19}
Jenkins,~A.; Pelliccione,~M.; Yu,~G.; Ma,~X.; Li,~X.; Wang,~K.~L.; Jayich,~A.
  C.~B. Single-spin sensing of domain-wall structure and dynamics in a
  thin-film skyrmion host. \emph{Phys. Rev. Materials} \textbf{2019}, \emph{3},
  083801\relax
\mciteBstWouldAddEndPuncttrue
\mciteSetBstMidEndSepPunct{\mcitedefaultmidpunct}
{\mcitedefaultendpunct}{\mcitedefaultseppunct}\relax
\EndOfBibitem
\bibitem[Thiel \latin{et~al.}(2016)Thiel, Rohner, Ganzhorn, Appel, Neu, Muller,
  Kleiner, Koelle, and Maletinsky]{thiel16}
Thiel,~L.; Rohner,~D.; Ganzhorn,~M.; Appel,~P.; Neu,~E.; Muller,~B.;
  Kleiner,~R.; Koelle,~D.; Maletinsky,~P. Quantitative nanoscale vortex imaging
  using a cryogenic quantum magnetometer. \emph{Nat. Nanotechnol.}
  \textbf{2016}, \emph{11}, 677\relax
\mciteBstWouldAddEndPuncttrue
\mciteSetBstMidEndSepPunct{\mcitedefaultmidpunct}
{\mcitedefaultendpunct}{\mcitedefaultseppunct}\relax
\EndOfBibitem
\bibitem[Pelliccione \latin{et~al.}(2016)Pelliccione, Jenkins, Ovartchaiyapong,
  Reetz, Emmanouilidou, Ni, and Jayich]{pelliccione16}
Pelliccione,~M.; Jenkins,~A.; Ovartchaiyapong,~P.; Reetz,~C.;
  Emmanouilidou,~E.; Ni,~N.; Jayich,~A. C.~B. Scanned probe imaging of
  nanoscale magnetism at cryogenic temperatures. \emph{Nat. Nanotechnol.}
  \textbf{2016}, \emph{11}, 700--705\relax
\mciteBstWouldAddEndPuncttrue
\mciteSetBstMidEndSepPunct{\mcitedefaultmidpunct}
{\mcitedefaultendpunct}{\mcitedefaultseppunct}\relax
\EndOfBibitem
\bibitem[Chang \latin{et~al.}(2017)Chang, Eichler, Rhensius, Lorenzelli, and
  Degen]{chang17}
Chang,~K.; Eichler,~A.; Rhensius,~J.; Lorenzelli,~L.; Degen,~C.~L. Nanoscale
  imaging of current density with a single-spin magnetometer. \emph{Nano
  Letters} \textbf{2017}, \emph{17}, 2367\relax
\mciteBstWouldAddEndPuncttrue
\mciteSetBstMidEndSepPunct{\mcitedefaultmidpunct}
{\mcitedefaultendpunct}{\mcitedefaultseppunct}\relax
\EndOfBibitem
\bibitem[Palm \latin{et~al.}(2024)Palm, Ding, Huxter, Taniguchi, Watanabe, and
  Degen]{palm24}
Palm,~M.~L.; Ding,~C.; Huxter,~W.~S.; Taniguchi,~T.; Watanabe,~K.; Degen,~C.~L.
  Observation of current whirlpools in graphene at room temperature.
  \emph{Science} \textbf{2024}, \emph{384}, 465--469\relax
\mciteBstWouldAddEndPuncttrue
\mciteSetBstMidEndSepPunct{\mcitedefaultmidpunct}
{\mcitedefaultendpunct}{\mcitedefaultseppunct}\relax
\EndOfBibitem
\bibitem[Schmid \latin{et~al.}(2010)Schmid, Marioni, Kappenberger, Romer,
  Parlinska-wojtan, Hug, Hellwig, Carey, and Fullerton]{schmid10}
Schmid,~I.; Marioni,~M.~A.; Kappenberger,~P.; Romer,~S.; Parlinska-wojtan,~M.;
  Hug,~H.~J.; Hellwig,~O.; Carey,~M.~J.; Fullerton,~E.~E. Exchange bias and
  domain evolution at 10 nm scales. \emph{Phys. Rev. Lett.} \textbf{2010},
  \emph{105}, 197201\relax
\mciteBstWouldAddEndPuncttrue
\mciteSetBstMidEndSepPunct{\mcitedefaultmidpunct}
{\mcitedefaultendpunct}{\mcitedefaultseppunct}\relax
\EndOfBibitem
\bibitem[Bode \latin{et~al.}(2006)Bode, Vedmedenko, Von~Bergmann, Kubetzka,
  Ferriani, Heinze, and Wiesendanger]{bode2006atomic}
Bode,~M.; Vedmedenko,~E.; Von~Bergmann,~K.; Kubetzka,~A.; Ferriani,~P.;
  Heinze,~S.; Wiesendanger,~R. Atomic spin structure of antiferromagnetic
  domain walls. \emph{Nature materials} \textbf{2006}, \emph{5}, 477--481\relax
\mciteBstWouldAddEndPuncttrue
\mciteSetBstMidEndSepPunct{\mcitedefaultmidpunct}
{\mcitedefaultendpunct}{\mcitedefaultseppunct}\relax
\EndOfBibitem
\bibitem[Zhang \latin{et~al.}(2017)Zhang, Chuu, Ren, Li, Li, Jin, Chou, and
  Shih]{zhang2017interlayer}
Zhang,~C.; Chuu,~C.-P.; Ren,~X.; Li,~M.-Y.; Li,~L.-J.; Jin,~C.; Chou,~M.-Y.;
  Shih,~C.-K. Interlayer couplings, Moir{\'e} patterns, and 2D electronic
  superlattices in MoS2/WSe2 hetero-bilayers. \emph{Science advances}
  \textbf{2017}, \emph{3}, e1601459\relax
\mciteBstWouldAddEndPuncttrue
\mciteSetBstMidEndSepPunct{\mcitedefaultmidpunct}
{\mcitedefaultendpunct}{\mcitedefaultseppunct}\relax
\EndOfBibitem
\bibitem[Crommie \latin{et~al.}(1993)Crommie, Lutz, and
  Eigler]{crommie1993confinement}
Crommie,~M.~F.; Lutz,~C.~P.; Eigler,~D.~M. Confinement of electrons to quantum
  corrals on a metal surface. \emph{Science} \textbf{1993}, \emph{262},
  218--220\relax
\mciteBstWouldAddEndPuncttrue
\mciteSetBstMidEndSepPunct{\mcitedefaultmidpunct}
{\mcitedefaultendpunct}{\mcitedefaultseppunct}\relax
\EndOfBibitem
\bibitem[Heller \latin{et~al.}(1994)Heller, Crommie, Lutz, and
  Eigler]{heller1994scattering}
Heller,~E.; Crommie,~M.; Lutz,~C.; Eigler,~D. Scattering and absorption of
  surface electron waves in quantum corrals. \emph{Nature} \textbf{1994},
  \emph{369}, 464--466\relax
\mciteBstWouldAddEndPuncttrue
\mciteSetBstMidEndSepPunct{\mcitedefaultmidpunct}
{\mcitedefaultendpunct}{\mcitedefaultseppunct}\relax
\EndOfBibitem
\bibitem[Tisler \latin{et~al.}(2013)Tisler, Oeckinghaus, Stöhr, Kolesov,
  Reuter, Reinhard, and Wrachtrup]{tisler2013single}
Tisler,~J.; Oeckinghaus,~T.; Stöhr,~R.~J.; Kolesov,~R.; Reuter,~R.;
  Reinhard,~F.; Wrachtrup,~J. Single defect center scanning near-field optical
  microscopy on graphene. \emph{Nano letters} \textbf{2013}, \emph{13},
  3152--3156\relax
\mciteBstWouldAddEndPuncttrue
\mciteSetBstMidEndSepPunct{\mcitedefaultmidpunct}
{\mcitedefaultendpunct}{\mcitedefaultseppunct}\relax
\EndOfBibitem
\bibitem[Rondin \latin{et~al.}(2014)Rondin, Tetienne, Hingant, Roch,
  Maletinsky, and Jacques]{rondin14}
Rondin,~L.; Tetienne,~J.~P.; Hingant,~T.; Roch,~J.~F.; Maletinsky,~P.;
  Jacques,~V. Magnetometry with nitrogen-vacancy defects in diamond. \emph{Rep.
  Prog. Phys.} \textbf{2014}, \emph{77}, 056503\relax
\mciteBstWouldAddEndPuncttrue
\mciteSetBstMidEndSepPunct{\mcitedefaultmidpunct}
{\mcitedefaultendpunct}{\mcitedefaultseppunct}\relax
\EndOfBibitem
\bibitem[Schirhagl \latin{et~al.}(2014)Schirhagl, Chang, Loretz, and
  Degen]{schirhagl14}
Schirhagl,~R.; Chang,~K.; Loretz,~M.; Degen,~C.~L. Nitrogen-vacancy centers in
  diamond: Nanoscale sensors for physics and biology. \emph{Annu. Rev. Phys.
  Chem.} \textbf{2014}, \emph{65}, 83\relax
\mciteBstWouldAddEndPuncttrue
\mciteSetBstMidEndSepPunct{\mcitedefaultmidpunct}
{\mcitedefaultendpunct}{\mcitedefaultseppunct}\relax
\EndOfBibitem
\bibitem[Maletinsky \latin{et~al.}(2012)Maletinsky, Hong, Grinolds, Hausmann,
  Lukin, Walsworth, Loncar, and Yacoby]{maletinsky12}
Maletinsky,~P.; Hong,~S.; Grinolds,~M.~S.; Hausmann,~B.; Lukin,~M.~D.;
  Walsworth,~R.~L.; Loncar,~M.; Yacoby,~A. A robust scanning diamond sensor for
  nanoscale imaging with single nitrogen-vacancy centres. \emph{Nat.
  Nanotechnol.} \textbf{2012}, \emph{7}, 320--324\relax
\mciteBstWouldAddEndPuncttrue
\mciteSetBstMidEndSepPunct{\mcitedefaultmidpunct}
{\mcitedefaultendpunct}{\mcitedefaultseppunct}\relax
\EndOfBibitem
\bibitem[Zhu \latin{et~al.}(2023)Zhu, Rhensius, Herb, Damle, Puebla-Hellmann,
  Degen, and Janitz]{zhu23}
Zhu,~T.; Rhensius,~J.; Herb,~K.; Damle,~V.; Puebla-Hellmann,~G.; Degen,~C.~L.;
  Janitz,~E. Multicone diamond waveguides for nanoscale quantum sensing.
  \emph{Nano Letters} \textbf{2023}, \emph{23}, 10110--10117\relax
\mciteBstWouldAddEndPuncttrue
\mciteSetBstMidEndSepPunct{\mcitedefaultmidpunct}
{\mcitedefaultendpunct}{\mcitedefaultseppunct}\relax
\EndOfBibitem
\bibitem[Hingant \latin{et~al.}(2015)Hingant, Tetienne, Mart{\'\i}nez, Garcia,
  Ravelosona, Roch, and Jacques]{hingant2015measuring}
Hingant,~T.; Tetienne,~J.-P.; Mart{\'\i}nez,~L.; Garcia,~K.; Ravelosona,~D.;
  Roch,~J.-F.; Jacques,~V. Measuring the magnetic moment density in patterned
  ultrathin ferromagnets with submicrometer resolution. \emph{Physical Review
  Applied} \textbf{2015}, \emph{4}, 014003\relax
\mciteBstWouldAddEndPuncttrue
\mciteSetBstMidEndSepPunct{\mcitedefaultmidpunct}
{\mcitedefaultendpunct}{\mcitedefaultseppunct}\relax
\EndOfBibitem
\bibitem[Tetienne \latin{et~al.}(2015)Tetienne, Hingant, Mart{\'\i}nez, Rohart,
  Thiaville, Diez, Garcia, Adam, Kim, Roch, \latin{et~al.}
  others]{tetienne2015nature}
Tetienne,~J.-P.; Hingant,~T.; Mart{\'\i}nez,~L.; Rohart,~S.; Thiaville,~A.;
  Diez,~L.~H.; Garcia,~K.; Adam,~J.-P.; Kim,~J.-V.; Roch,~J.-F.; others The
  nature of domain walls in ultrathin ferromagnets revealed by scanning
  nanomagnetometry. \emph{Nature communications} \textbf{2015}, \emph{6},
  1--6\relax
\mciteBstWouldAddEndPuncttrue
\mciteSetBstMidEndSepPunct{\mcitedefaultmidpunct}
{\mcitedefaultendpunct}{\mcitedefaultseppunct}\relax
\EndOfBibitem
\bibitem[Gross \latin{et~al.}(2016)Gross, Mart{\'\i}nez, Tetienne, Hingant,
  Roch, Garcia, Soucaille, Adam, Kim, Rohart, \latin{et~al.}
  others]{gross2016direct}
Gross,~I.; Mart{\'\i}nez,~L.; Tetienne,~J.-P.; Hingant,~T.; Roch,~J.-F.;
  Garcia,~K.; Soucaille,~R.; Adam,~J.; Kim,~J.-V.; Rohart,~S.; others Direct
  measurement of interfacial Dzyaloshinskii-Moriya interaction in X| CoFeB| MgO
  heterostructures with a scanning NV magnetometer (X= Ta, TaN, and W).
  \emph{Physical Review B} \textbf{2016}, \emph{94}, 064413\relax
\mciteBstWouldAddEndPuncttrue
\mciteSetBstMidEndSepPunct{\mcitedefaultmidpunct}
{\mcitedefaultendpunct}{\mcitedefaultseppunct}\relax
\EndOfBibitem
\bibitem[Rohner \latin{et~al.}(2019)Rohner, Happacher, Reiser, Tschudin,
  Tallaire, Achard, Shields, and Maletinsky]{rohner2019111}
Rohner,~D.; Happacher,~J.; Reiser,~P.; Tschudin,~M.; Tallaire,~A.; Achard,~J.;
  Shields,~B.; Maletinsky,~P. (111)-oriented, single crystal diamond tips for
  nanoscale scanning probe imaging of out-of-plane magnetic fields.
  \emph{Applied Physics Letters} \textbf{2019}, \emph{115}, 192401\relax
\mciteBstWouldAddEndPuncttrue
\mciteSetBstMidEndSepPunct{\mcitedefaultmidpunct}
{\mcitedefaultendpunct}{\mcitedefaultseppunct}\relax
\EndOfBibitem
\bibitem[Zhong \latin{et~al.}(2022)Zhong, Finco, Fischer, Haykal, Bouzehouane,
  Carr{\'e}t{\'e}ro, Godel, Maletinsky, Munsch, Fusil, \latin{et~al.}
  others]{zhong2022quantitative}
Zhong,~H.; Finco,~A.; Fischer,~J.; Haykal,~A.; Bouzehouane,~K.;
  Carr{\'e}t{\'e}ro,~C.; Godel,~F.; Maletinsky,~P.; Munsch,~M.; Fusil,~S.;
  others Quantitative Imaging of Exotic Antiferromagnetic Spin Cycloids in Bi
  Fe O 3 Thin Films. \emph{Physical Review Applied} \textbf{2022}, \emph{17},
  044051\relax
\mciteBstWouldAddEndPuncttrue
\mciteSetBstMidEndSepPunct{\mcitedefaultmidpunct}
{\mcitedefaultendpunct}{\mcitedefaultseppunct}\relax
\EndOfBibitem
\bibitem[Finco \latin{et~al.}(2022)Finco, Haykal, Fusil, Kumar, Dufour, Forget,
  Colson, Chauleau, Viret, Jaouen, \latin{et~al.} others]{finco2022imaging}
Finco,~A.; Haykal,~A.; Fusil,~S.; Kumar,~P.; Dufour,~P.; Forget,~A.;
  Colson,~D.; Chauleau,~J.-Y.; Viret,~M.; Jaouen,~N.; others Imaging
  topological defects in a noncollinear antiferromagnet. \emph{Physical Review
  Letters} \textbf{2022}, \emph{128}, 187201\relax
\mciteBstWouldAddEndPuncttrue
\mciteSetBstMidEndSepPunct{\mcitedefaultmidpunct}
{\mcitedefaultendpunct}{\mcitedefaultseppunct}\relax
\EndOfBibitem
\bibitem[Finco and Jacques(2023)Finco, and Jacques]{finco2023single}
Finco,~A.; Jacques,~V. Single spin magnetometry and relaxometry applied to
  antiferromagnetic materials. \emph{APL Materials} \textbf{2023},
  \emph{11}\relax
\mciteBstWouldAddEndPuncttrue
\mciteSetBstMidEndSepPunct{\mcitedefaultmidpunct}
{\mcitedefaultendpunct}{\mcitedefaultseppunct}\relax
\EndOfBibitem
\bibitem[Pham \latin{et~al.}(2024)Pham, Sisodia, di~Manici,
  Urrestarazu-larranaga, Bairagi, Pelloux-Prayer, Guedas, Buda-Prejbeanu,
  Auffret, Locatelli, Mentes, Pizzini, Kumar, Finco, Jacques, Gaudin, and
  Boulle]{pham24}
Pham,~V.~T. \latin{et~al.}  Fast current-induced skyrmion motion in synthetic
  antiferromagnets. \emph{Science} \textbf{2024}, \emph{384}, 307--312\relax
\mciteBstWouldAddEndPuncttrue
\mciteSetBstMidEndSepPunct{\mcitedefaultmidpunct}
{\mcitedefaultendpunct}{\mcitedefaultseppunct}\relax
\EndOfBibitem
\bibitem[{QZabre Ltd.}()]{qzabre}
{QZabre Ltd.} {https://qzabre.com}. \relax
\mciteBstWouldAddEndPunctfalse
\mciteSetBstMidEndSepPunct{\mcitedefaultmidpunct}
{}{\mcitedefaultseppunct}\relax
\EndOfBibitem
\bibitem[Janitz \latin{et~al.}(2022)Janitz, Herb, Volker, Huxter, Degen, and
  Abendroth]{janitz22}
Janitz,~E.; Herb,~K.; Volker,~L.~A.; Huxter,~W.~S.; Degen,~C.~L.;
  Abendroth,~J.~M. Diamond surface engineering for molecular sensing with
  nitrogen-vacancy centers. \emph{Journal of Materials Chemistry C}
  \textbf{2022}, \emph{10}, 13533--13569\relax
\mciteBstWouldAddEndPuncttrue
\mciteSetBstMidEndSepPunct{\mcitedefaultmidpunct}
{\mcitedefaultendpunct}{\mcitedefaultseppunct}\relax
\EndOfBibitem
\bibitem[Lo \latin{et~al.}(1999)Lo, Huefner, Chan, Dryden, Hagenhoff, and
  Beebe]{lo1999}
Lo,~Y.; Huefner,~N.~D.; Chan,~W.~S.; Dryden,~P.; Hagenhoff,~B.; Beebe,~T.~P.
  Organic and inorganic contamination on commercial afm cantilevers.
  \emph{Langmuir} \textbf{1999}, \emph{15}, 6522--6526\relax
\mciteBstWouldAddEndPuncttrue
\mciteSetBstMidEndSepPunct{\mcitedefaultmidpunct}
{\mcitedefaultendpunct}{\mcitedefaultseppunct}\relax
\EndOfBibitem
\bibitem[Parthasarathy \latin{et~al.}(2024)Parthasarathy, Joos, Hughes,
  Meynell, Morrison, Risner-Jamtgaard, Weld, Mukherjee, and
  Jayich]{parthasarathy2024role}
Parthasarathy,~S.; Joos,~M.; Hughes,~L.~B.; Meynell,~S.~A.; Morrison,~T.~A.;
  Risner-Jamtgaard,~J.; Weld,~D.~M.; Mukherjee,~K.; Jayich,~A. C.~B. Role of
  Oxygen in Laser Induced Contamination at Diamond-Vacuum Interfaces.
  \emph{arXiv preprint arXiv:2401.06942} \textbf{2024}, \relax
\mciteBstWouldAddEndPunctfalse
\mciteSetBstMidEndSepPunct{\mcitedefaultmidpunct}
{}{\mcitedefaultseppunct}\relax
\EndOfBibitem
\bibitem[Enachescu \latin{et~al.}(2004)Enachescu, Carpick, Ogletree, and
  Salmeron]{enachescu2004}
Enachescu,~M.; Carpick,~R.~W.; Ogletree,~D.~F.; Salmeron,~M. The role of
  contaminants in the variation of adhesion friction and electrical conduction
  properties of carbide-coated scanning probe tips and Pt(111) in ultrahigh
  vacuum. \emph{Journal of Applied Physics} \textbf{2004}, \emph{95},
  7694\relax
\mciteBstWouldAddEndPuncttrue
\mciteSetBstMidEndSepPunct{\mcitedefaultmidpunct}
{\mcitedefaultendpunct}{\mcitedefaultseppunct}\relax
\EndOfBibitem
\bibitem[Hoppe \latin{et~al.}(2005)Hoppe, Ctistis, Paggel, and
  Fumagalli]{hoppe2005}
Hoppe,~S.; Ctistis,~G.; Paggel,~J.~J.; Fumagalli,~P. Spectroscopy of the shear
  force interaction in scanning near-field optical microscopy.
  \emph{Ultramicroscopy} \textbf{2005}, \emph{102}, 221--226\relax
\mciteBstWouldAddEndPuncttrue
\mciteSetBstMidEndSepPunct{\mcitedefaultmidpunct}
{\mcitedefaultendpunct}{\mcitedefaultseppunct}\relax
\EndOfBibitem
\bibitem[Buchler \latin{et~al.}(2005)Buchler, Kalkbrenner, Hettich, and
  Sandoghdar]{buchler2005measuring}
Buchler,~B.; Kalkbrenner,~T.; Hettich,~C.; Sandoghdar,~V. Measuring the quantum
  efficiency of the optical emission of single radiating dipoles using a
  scanning mirror. \emph{Physical review letters} \textbf{2005}, \emph{95},
  063003\relax
\mciteBstWouldAddEndPuncttrue
\mciteSetBstMidEndSepPunct{\mcitedefaultmidpunct}
{\mcitedefaultendpunct}{\mcitedefaultseppunct}\relax
\EndOfBibitem
\bibitem[Israelsen \latin{et~al.}(2014)Israelsen, Kumar, Tawfieq,
  Neergaard-Nielsen, Huck, and Andersen]{israelsen2014increasing}
Israelsen,~N.~M.; Kumar,~S.; Tawfieq,~M.; Neergaard-Nielsen,~J.~S.; Huck,~A.;
  Andersen,~U.~L. Increasing the photon collection rate from a single NV center
  with a silver mirror. \emph{Journal of optics} \textbf{2014}, \emph{16},
  114017\relax
\mciteBstWouldAddEndPuncttrue
\mciteSetBstMidEndSepPunct{\mcitedefaultmidpunct}
{\mcitedefaultendpunct}{\mcitedefaultseppunct}\relax
\EndOfBibitem
\bibitem[Ernst \latin{et~al.}(2019)Ernst, Irber, Waeber, Braunbeck, and
  Reinhard]{ernst2019planar}
Ernst,~S.; Irber,~D.~M.; Waeber,~A.~M.; Braunbeck,~G.; Reinhard,~F. A planar
  scanning probe microscope. \emph{ACS Photonics} \textbf{2019}, \emph{6},
  327--331\relax
\mciteBstWouldAddEndPuncttrue
\mciteSetBstMidEndSepPunct{\mcitedefaultmidpunct}
{\mcitedefaultendpunct}{\mcitedefaultseppunct}\relax
\EndOfBibitem
\bibitem[Drummond~Roby and Wetsel~Jr(1996)Drummond~Roby, and
  Wetsel~Jr]{drummond1996measurement}
Drummond~Roby,~M.; Wetsel~Jr,~G. Measurement of elastic force on a scanned
  probe near a solid surface. \emph{Applied physics letters} \textbf{1996},
  \emph{69}, 3689--3691\relax
\mciteBstWouldAddEndPuncttrue
\mciteSetBstMidEndSepPunct{\mcitedefaultmidpunct}
{\mcitedefaultendpunct}{\mcitedefaultseppunct}\relax
\EndOfBibitem
\bibitem[Pfeiffer \latin{et~al.}(2002)Pfeiffer, Bennewitz, Baratoff, Meyer, and
  Gr{\"u}tter]{pfeiffer2002lateral}
Pfeiffer,~O.; Bennewitz,~R.; Baratoff,~A.; Meyer,~E.; Gr{\"u}tter,~P.
  Lateral-force measurements in dynamic force microscopy. \emph{Physical review
  B} \textbf{2002}, \emph{65}, 161403\relax
\mciteBstWouldAddEndPuncttrue
\mciteSetBstMidEndSepPunct{\mcitedefaultmidpunct}
{\mcitedefaultendpunct}{\mcitedefaultseppunct}\relax
\EndOfBibitem
\bibitem[Karrai and Tiemann(2000)Karrai, and Tiemann]{karrai2000interfacial}
Karrai,~K.; Tiemann,~I. Interfacial shear force microscopy. \emph{Physical
  Review B} \textbf{2000}, \emph{62}, 13174\relax
\mciteBstWouldAddEndPuncttrue
\mciteSetBstMidEndSepPunct{\mcitedefaultmidpunct}
{\mcitedefaultendpunct}{\mcitedefaultseppunct}\relax
\EndOfBibitem
\bibitem[G{\"o}ttlich \latin{et~al.}(2000)G{\"o}ttlich, Stark, Pedarnig, and
  Heckl]{gottlich2000noncontact}
G{\"o}ttlich,~H.; Stark,~R.~W.; Pedarnig,~J.~D.; Heckl,~W.~M. Noncontact
  scanning force microscopy based on a modified tuning fork sensor.
  \emph{Review of Scientific Instruments} \textbf{2000}, \emph{71},
  3104--3107\relax
\mciteBstWouldAddEndPuncttrue
\mciteSetBstMidEndSepPunct{\mcitedefaultmidpunct}
{\mcitedefaultendpunct}{\mcitedefaultseppunct}\relax
\EndOfBibitem
\bibitem[W{\"o}rnle(2021)]{wornle2021nanoscale}
W{\"o}rnle,~M.~S. Nanoscale scanning diamond magnetometry of antiferromagnets.
  Ph.D.\ thesis, ETH Zurich, 2021\relax
\mciteBstWouldAddEndPuncttrue
\mciteSetBstMidEndSepPunct{\mcitedefaultmidpunct}
{\mcitedefaultendpunct}{\mcitedefaultseppunct}\relax
\EndOfBibitem
\bibitem[Luo \latin{et~al.}(2019)Luo, Dao, Hrabec, Vijayakumar, Kleibert,
  Baumgartner, Kirk, Cui, Savchenko, Krishnaswamy, \latin{et~al.}
  others]{luo2019chirally}
Luo,~Z.; Dao,~T.~P.; Hrabec,~A.; Vijayakumar,~J.; Kleibert,~A.;
  Baumgartner,~M.; Kirk,~E.; Cui,~J.; Savchenko,~T.; Krishnaswamy,~G.; others
  Chirally coupled nanomagnets. \emph{Science} \textbf{2019}, \emph{363},
  1435--1439\relax
\mciteBstWouldAddEndPuncttrue
\mciteSetBstMidEndSepPunct{\mcitedefaultmidpunct}
{\mcitedefaultendpunct}{\mcitedefaultseppunct}\relax
\EndOfBibitem
\bibitem[Mamin \latin{et~al.}(2013)Mamin, Kim, Sherwood, Rettner, Ohno,
  Awschalom, and Rugar]{mamin13}
Mamin,~H.~J.; Kim,~M.; Sherwood,~M.~H.; Rettner,~C.~T.; Ohno,~K.;
  Awschalom,~D.~D.; Rugar,~D. Nanoscale nuclear magnetic resonance with a
  nitrogen-vacancy spin sensor. \emph{Science} \textbf{2013}, \emph{339},
  557--560\relax
\mciteBstWouldAddEndPuncttrue
\mciteSetBstMidEndSepPunct{\mcitedefaultmidpunct}
{\mcitedefaultendpunct}{\mcitedefaultseppunct}\relax
\EndOfBibitem
\bibitem[Staudacher \latin{et~al.}(2013)Staudacher, Shi, Pezzagna, Meijer, Du,
  Meriles, Reinhard, and Wrachtrup]{staudacher13}
Staudacher,~T.; Shi,~F.; Pezzagna,~S.; Meijer,~J.; Du,~J.; Meriles,~C.~A.;
  Reinhard,~F.; Wrachtrup,~J. Nuclear magnetic resonance spectroscopy on a
  (5-nanometer)$^3$ sample volume. \emph{Science} \textbf{2013}, \emph{339},
  561--563\relax
\mciteBstWouldAddEndPuncttrue
\mciteSetBstMidEndSepPunct{\mcitedefaultmidpunct}
{\mcitedefaultendpunct}{\mcitedefaultseppunct}\relax
\EndOfBibitem
\bibitem[Degen \latin{et~al.}(2009)Degen, Poggio, Mamin, Rettner, and
  Rugar]{degen09}
Degen,~C.~L.; Poggio,~M.; Mamin,~H.~J.; Rettner,~C.~T.; Rugar,~D. Nanoscale
  magnetic resonance imaging. \emph{Proc. Nat. Acad. Sci. U.S.A.}
  \textbf{2009}, \emph{106}, 1313\relax
\mciteBstWouldAddEndPuncttrue
\mciteSetBstMidEndSepPunct{\mcitedefaultmidpunct}
{\mcitedefaultendpunct}{\mcitedefaultseppunct}\relax
\EndOfBibitem
\bibitem[Loretz \latin{et~al.}(2014)Loretz, Pezzagna, Meijer, and
  Degen]{loretz14apl}
Loretz,~M.; Pezzagna,~S.; Meijer,~J.; Degen,~C.~L. Nanoscale nuclear magnetic
  resonance with a 1.9-nm-deep nitrogen-vacancy sensor. \emph{Appl. Phys.
  Lett.} \textbf{2014}, \emph{104}, 033102\relax
\mciteBstWouldAddEndPuncttrue
\mciteSetBstMidEndSepPunct{\mcitedefaultmidpunct}
{\mcitedefaultendpunct}{\mcitedefaultseppunct}\relax
\EndOfBibitem
\bibitem[Pham \latin{et~al.}(2016)Pham, Devience, Casola, Lovchinsky, Sushkov,
  Bersin, Lee, Urbach, Cappellaro, Park, Yacoby, Lukin, and Walsworth]{pham16}
Pham,~L.~M.; Devience,~S.~J.; Casola,~F.; Lovchinsky,~I.; Sushkov,~A.~O.;
  Bersin,~E.; Lee,~J.; Urbach,~E.; Cappellaro,~P.; Park,~H.; Yacoby,~A.;
  Lukin,~M.; Walsworth,~R.~L. {{NMR}} technique for determining the depth of
  shallow nitrogen-vacancy centers in diamond. \emph{Phys. Rev. B}
  \textbf{2016}, \emph{93}, 045425\relax
\mciteBstWouldAddEndPuncttrue
\mciteSetBstMidEndSepPunct{\mcitedefaultmidpunct}
{\mcitedefaultendpunct}{\mcitedefaultseppunct}\relax
\EndOfBibitem
\bibitem[Bylander \latin{et~al.}(2011)Bylander, Gustavsson, Yan, Yoshihara,
  Harrabi, Fitch, Cory, Nakamura, Tsai, and Oliver]{bylander11}
Bylander,~J.; Gustavsson,~S.; Yan,~F.; Yoshihara,~F.; Harrabi,~K.; Fitch,~G.;
  Cory,~D.~G.; Nakamura,~Y.; Tsai,~J.~S.; Oliver,~W.~D. Noise spectroscopy
  through dynamical decoupling with a superconducting flux qubit. \emph{Nat.
  Phys.} \textbf{2011}, \emph{7}, 565--570\relax
\mciteBstWouldAddEndPuncttrue
\mciteSetBstMidEndSepPunct{\mcitedefaultmidpunct}
{\mcitedefaultendpunct}{\mcitedefaultseppunct}\relax
\EndOfBibitem
\bibitem[Ziegler and Biersack(1985)Ziegler, and Biersack]{ziegler1985stopping}
Ziegler,~J.~F.; Biersack,~J.~P. \emph{Treatise on heavy-ion science: volume 6:
  astrophysics, chemistry, and condensed matter}; Springer, 1985; pp
  93--129\relax
\mciteBstWouldAddEndPuncttrue
\mciteSetBstMidEndSepPunct{\mcitedefaultmidpunct}
{\mcitedefaultendpunct}{\mcitedefaultseppunct}\relax
\EndOfBibitem
\bibitem[Abendroth \latin{et~al.}(2022)Abendroth, Herb, Janitz, Zhu, Volker,
  and Degen]{abendroth22}
Abendroth,~J.~M.; Herb,~K.; Janitz,~E.; Zhu,~T.; Volker,~L.~A.; Degen,~C.~L.
  Single-nitrogen-vacancy {NMR} of amine-functionalized diamond surfaces.
  \emph{Nano Letters} \textbf{2022}, \emph{22}, 7294--7303\relax
\mciteBstWouldAddEndPuncttrue
\mciteSetBstMidEndSepPunct{\mcitedefaultmidpunct}
{\mcitedefaultendpunct}{\mcitedefaultseppunct}\relax
\EndOfBibitem
\bibitem[Weeks \latin{et~al.}(2005)Weeks, Vaughn, and Deyoreo]{weeks2005}
Weeks,~B.~L.; Vaughn,~M.~W.; Deyoreo,~J.~J. Direct imaging of meniscus
  formation in atomic force microscopy using environmental scanning electron
  microscopy. \emph{Langmuir} \textbf{2005}, \emph{21}, 8096--8098\relax
\mciteBstWouldAddEndPuncttrue
\mciteSetBstMidEndSepPunct{\mcitedefaultmidpunct}
{\mcitedefaultendpunct}{\mcitedefaultseppunct}\relax
\EndOfBibitem
\bibitem[Gan and Franks(2009)Gan, and Franks]{gan2009}
Gan,~Y.; Franks,~G.~V. Cleaning AFM colloidal probes by mechanically scrubbing
  with supersharp brushes. \emph{Ultramicroscopy} \textbf{2009}, \emph{109},
  1061--1065\relax
\mciteBstWouldAddEndPuncttrue
\mciteSetBstMidEndSepPunct{\mcitedefaultmidpunct}
{\mcitedefaultendpunct}{\mcitedefaultseppunct}\relax
\EndOfBibitem
\bibitem[Chauleau \latin{et~al.}(2020)Chauleau, Chirac, Fusil, Garcia, Akhtar,
  Tranchida, Thibaudeau, Gross, Blouzon, Finco, \latin{et~al.}
  others]{chauleau2020electric}
Chauleau,~J.-Y.; Chirac,~T.; Fusil,~S.; Garcia,~V.; Akhtar,~W.; Tranchida,~J.;
  Thibaudeau,~P.; Gross,~I.; Blouzon,~C.; Finco,~A.~e.; others Electric and
  antiferromagnetic chiral textures at multiferroic domain walls. \emph{Nature
  Materials} \textbf{2020}, \emph{19}, 386--390\relax
\mciteBstWouldAddEndPuncttrue
\mciteSetBstMidEndSepPunct{\mcitedefaultmidpunct}
{\mcitedefaultendpunct}{\mcitedefaultseppunct}\relax
\EndOfBibitem
\bibitem[Haykal \latin{et~al.}(2020)Haykal, Fischer, Akhtar, Chauleau, Sando,
  Finco, Godel, Birkh{\"o}lzer, Carr{\'e}t{\'e}ro, Jaouen, \latin{et~al.}
  others]{haykal2020antiferromagnetic}
Haykal,~A.; Fischer,~J.; Akhtar,~W.; Chauleau,~J.-Y.; Sando,~D.; Finco,~A.;
  Godel,~F.; Birkh{\"o}lzer,~Y.; Carr{\'e}t{\'e}ro,~C.; Jaouen,~N.; others
  Antiferromagnetic textures in BiFeO3 controlled by strain and electric field.
  \emph{Nature communications} \textbf{2020}, \emph{11}, 1--7\relax
\mciteBstWouldAddEndPuncttrue
\mciteSetBstMidEndSepPunct{\mcitedefaultmidpunct}
{\mcitedefaultendpunct}{\mcitedefaultseppunct}\relax
\EndOfBibitem
\bibitem[Dufour \latin{et~al.}(2023)Dufour, Abdelsamie, Fischer, Finco, Haykal,
  Sarott, Varotto, Carr{\'e}t{\'e}ro, Collin, Godel, \latin{et~al.}
  others]{dufour2023onset}
Dufour,~P.; Abdelsamie,~A.; Fischer,~J.; Finco,~A.; Haykal,~A.; Sarott,~M.~F.;
  Varotto,~S.; Carr{\'e}t{\'e}ro,~C.; Collin,~S.; Godel,~F.; others Onset of
  multiferroicity in prototypical single-spin cycloid BiFeO3 thin films.
  \emph{Nano Letters} \textbf{2023}, \emph{23}, 9073--9079\relax
\mciteBstWouldAddEndPuncttrue
\mciteSetBstMidEndSepPunct{\mcitedefaultmidpunct}
{\mcitedefaultendpunct}{\mcitedefaultseppunct}\relax
\EndOfBibitem
\bibitem[Rugar \latin{et~al.}(2015)Rugar, Mamin, Sherwood, Kim, Rettner, Ohno,
  and Awschalom]{rugar2015proton}
Rugar,~D.; Mamin,~H.; Sherwood,~M.; Kim,~M.; Rettner,~C.~T.; Ohno,~K.;
  Awschalom,~D.~D. Proton magnetic resonance imaging using a nitrogen--vacancy
  spin sensor. \emph{Nature nanotechnology} \textbf{2015}, \emph{10},
  120--124\relax
\mciteBstWouldAddEndPuncttrue
\mciteSetBstMidEndSepPunct{\mcitedefaultmidpunct}
{\mcitedefaultendpunct}{\mcitedefaultseppunct}\relax
\EndOfBibitem
\bibitem[Ku \latin{et~al.}(2020)Ku, Zhou, Li, Shin, Shi, Burch, Anderson,
  Pierce, Xie, Hamo, Vool, Zhang, Casola, Taniguchi, Watanabe, Fogler, Kim,
  Yacoby, and Walsworth]{ku20}
Ku,~M. J.~H. \latin{et~al.}  Imaging viscous flow of the {{Dirac}} fluid in
  graphene. \emph{Nature} \textbf{2020}, \emph{583}, 537--541\relax
\mciteBstWouldAddEndPuncttrue
\mciteSetBstMidEndSepPunct{\mcitedefaultmidpunct}
{\mcitedefaultendpunct}{\mcitedefaultseppunct}\relax
\EndOfBibitem
\bibitem[Palm \latin{et~al.}(2022)Palm, Huxter, Welter, Ernst, Scheidegger,
  Diesch, Chang, Rickhaus, Taniguchi, Watanabe, Ensslin, and Degen]{palm22}
Palm,~M.~L.; Huxter,~W.~S.; Welter,~P.; Ernst,~S.; Scheidegger,~P.~J.;
  Diesch,~S.; Chang,~K.; Rickhaus,~P.; Taniguchi,~T.; Watanabe,~K.;
  Ensslin,~K.; Degen,~C.~L. Imaging of Submicroampere Currents in Bilayer
  Graphene Using a Scanning Diamond Magnetometer. \emph{Physical Review
  Applied} \textbf{2022}, \emph{17}\relax
\mciteBstWouldAddEndPuncttrue
\mciteSetBstMidEndSepPunct{\mcitedefaultmidpunct}
{\mcitedefaultendpunct}{\mcitedefaultseppunct}\relax
\EndOfBibitem
\bibitem[Huxter \latin{et~al.}(2022)Huxter, Palm, Davis, Welter, Lambert,
  Trassin, and Degen]{huxter22}
Huxter,~W.~S.; Palm,~M.~L.; Davis,~M.~L.; Welter,~P.; Lambert,~C.~H.;
  Trassin,~M.; Degen,~C.~L. Scanning gradiometry with a single spin quantum
  magnetometer. \emph{Nature Communications} \textbf{2022}, \emph{13},
  3761\relax
\mciteBstWouldAddEndPuncttrue
\mciteSetBstMidEndSepPunct{\mcitedefaultmidpunct}
{\mcitedefaultendpunct}{\mcitedefaultseppunct}\relax
\EndOfBibitem
\bibitem[Huxter \latin{et~al.}(2023)Huxter, Sarott, Trassin, and
  Degen]{huxter23}
Huxter,~W.~S.; Sarott,~M.~F.; Trassin,~M.; Degen,~C.~L. Imaging ferroelectric
  domains with a single-spin scanning quantum sensor. \emph{Nature Physics}
  \textbf{2023}, \emph{19}, 644\relax
\mciteBstWouldAddEndPuncttrue
\mciteSetBstMidEndSepPunct{\mcitedefaultmidpunct}
{\mcitedefaultendpunct}{\mcitedefaultseppunct}\relax
\EndOfBibitem
\bibitem[Sangtawesin \latin{et~al.}(2019)Sangtawesin, Dwyer, Srinivasan,
  Allred, Rodgers, Greve, Stacey, Dontschuk, ODonnell, Hu, Evans, Jaye,
  Fischer, Markham, Twitchen, Park, Lukin, and de~Leon]{sangtawesin19}
Sangtawesin,~S. \latin{et~al.}  Origins of diamond surface noise probed by
  correlating single-spin measurements with surface spectroscopy. \emph{Phys.
  Rev. X} \textbf{2019}, \emph{9}, 031052\relax
\mciteBstWouldAddEndPuncttrue
\mciteSetBstMidEndSepPunct{\mcitedefaultmidpunct}
{\mcitedefaultendpunct}{\mcitedefaultseppunct}\relax
\EndOfBibitem
\bibitem[Ohashi \latin{et~al.}(2013)Ohashi, Rosskopf, Watanabe, Loretz, Tao,
  Hauert, Tomizawa, Ishikawa, Ishi-hayase, Shikata, Degen, and Itoh]{ohashi13}
Ohashi,~K.; Rosskopf,~T.; Watanabe,~H.; Loretz,~M.; Tao,~Y.; Hauert,~R.;
  Tomizawa,~S.; Ishikawa,~T.; Ishi-hayase,~J.; Shikata,~S.; Degen,~C.~L.;
  Itoh,~K.~M. Negatively charged nitrogen-vacancy centers in a 5 nm thin 12C
  diamond film. \emph{Nano Letters} \textbf{2013}, \emph{13}, 4733--4738\relax
\mciteBstWouldAddEndPuncttrue
\mciteSetBstMidEndSepPunct{\mcitedefaultmidpunct}
{\mcitedefaultendpunct}{\mcitedefaultseppunct}\relax
\EndOfBibitem
\bibitem[Myers \latin{et~al.}(2014)Myers, Das, Dartiailh, Ohno, Awschalom, and
  Jayich]{myers14}
Myers,~B.~A.; Das,~A.; Dartiailh,~M.~C.; Ohno,~K.; Awschalom,~D.~D.; Jayich,~A.
  C.~B. Probing surface noise with depth-calibrated spins in diamond.
  \emph{Phys. Rev. Lett.} \textbf{2014}, \emph{113}, 027602\relax
\mciteBstWouldAddEndPuncttrue
\mciteSetBstMidEndSepPunct{\mcitedefaultmidpunct}
{\mcitedefaultendpunct}{\mcitedefaultseppunct}\relax
\EndOfBibitem
\bibitem[Momenzadeh \latin{et~al.}(2015)Momenzadeh, Stohr, oliveira, Brunner,
  Denisenko, Yang, Reinhard, and Wrachtrup]{momenzadeh15}
Momenzadeh,~S.~A.; Stohr,~R.~J.; oliveira,~F. F.~D.; Brunner,~A.;
  Denisenko,~A.; Yang,~S.; Reinhard,~F.; Wrachtrup,~J. Nanoengineered diamond
  waveguide as a robust bright platform for nanomagnetometry using shallow
  nitrogen vacancy centers. \emph{Nano Letters} \textbf{2015}, \emph{15},
  165--169\relax
\mciteBstWouldAddEndPuncttrue
\mciteSetBstMidEndSepPunct{\mcitedefaultmidpunct}
{\mcitedefaultendpunct}{\mcitedefaultseppunct}\relax
\EndOfBibitem
\bibitem[Hedrich \latin{et~al.}(2020)Hedrich, Rohner, Batzer, Maletinsky, and
  Shields]{hedrich20}
Hedrich,~N.; Rohner,~D.; Batzer,~M.; Maletinsky,~P.; Shields,~B.~J. Parabolic
  diamond scanning probes for single-spin magnetic field imaging.
  \emph{Physical Review Applied} \textbf{2020}, \emph{14}, 064007\relax
\mciteBstWouldAddEndPuncttrue
\mciteSetBstMidEndSepPunct{\mcitedefaultmidpunct}
{\mcitedefaultendpunct}{\mcitedefaultseppunct}\relax
\EndOfBibitem
\bibitem[Jones \latin{et~al.}(2023)Jones, Delegan, Heremans, and
  Martinson]{jones23}
Jones,~J.~C.; Delegan,~N.; Heremans,~F.~J.; Martinson,~A.~B. Nucleation
  dependence of atomic layer deposition on diamond surface termination.
  \emph{Carbon} \textbf{2023}, \emph{213}, 118276\relax
\mciteBstWouldAddEndPuncttrue
\mciteSetBstMidEndSepPunct{\mcitedefaultmidpunct}
{\mcitedefaultendpunct}{\mcitedefaultseppunct}\relax
\EndOfBibitem
\bibitem[Edwards \latin{et~al.}(1997)Edwards, Taylor, Duncan, and
  Melmed]{edwards1997fast}
Edwards,~H.; Taylor,~L.; Duncan,~W.; Melmed,~A.~J. Fast, high-resolution atomic
  force microscopy using a quartz tuning fork as actuator and sensor.
  \emph{Journal of applied physics} \textbf{1997}, \emph{82}, 980--984\relax
\mciteBstWouldAddEndPuncttrue
\mciteSetBstMidEndSepPunct{\mcitedefaultmidpunct}
{\mcitedefaultendpunct}{\mcitedefaultseppunct}\relax
\EndOfBibitem
\bibitem[Giessibl(1998)]{giessibl1998high}
Giessibl,~F.~J. High-speed force sensor for force microscopy and profilometry
  utilizing a quartz tuning fork. \emph{Applied physics letters} \textbf{1998},
  \emph{73}, 3956--3958\relax
\mciteBstWouldAddEndPuncttrue
\mciteSetBstMidEndSepPunct{\mcitedefaultmidpunct}
{\mcitedefaultendpunct}{\mcitedefaultseppunct}\relax
\EndOfBibitem
\bibitem[Ruiter \latin{et~al.}(1998)Ruiter, Van Der~Werf, Veerman,
  Garcia-Parajo, Rensen, and Van~Hulst]{ruiter1998tuning}
Ruiter,~A.; Van Der~Werf,~K.; Veerman,~J.; Garcia-Parajo,~M.; Rensen,~W.;
  Van~Hulst,~N. Tuning fork shear-force feedback. \emph{Ultramicroscopy}
  \textbf{1998}, \emph{71}, 149--157\relax
\mciteBstWouldAddEndPuncttrue
\mciteSetBstMidEndSepPunct{\mcitedefaultmidpunct}
{\mcitedefaultendpunct}{\mcitedefaultseppunct}\relax
\EndOfBibitem
\end{mcitethebibliography}
%\input{"references.bbl"}
\providecommand{\noopsort}[1]{}\providecommand{\singleletter}[1]{#1}%
\providecommand{\latin}[1]{#1}
\makeatletter
\providecommand{\doi}
  {\begingroup\let\do\@makeother\dospecials
  \catcode`\{=1 \catcode`\}=2 \doi@aux}
\providecommand{\doi@aux}[1]{\endgroup\texttt{#1}}
\makeatother
\providecommand*\mcitethebibliography{\thebibliography}
\csname @ifundefined\endcsname{endmcitethebibliography}
  {\let\endmcitethebibliography\endthebibliography}{}

\end{document}